\documentclass[iop,revtex4]{emulateapj}

\usepackage{natbib}
\usepackage{graphicx}
\usepackage[space]{grffile}
\usepackage{latexsym}
\usepackage{amsfonts,amsmath,amssymb}
\usepackage{url}
\usepackage[utf8]{inputenc}
\usepackage{fancyref}
\usepackage{hyperref}
\hypersetup{colorlinks=false,pdfborder={0 0 0},}
\usepackage{textcomp}
\usepackage{longtable}
\usepackage{multirow,booktabs}
\usepackage{pdflscape}

\shorttitle{Fundamental Parameters of Brown Dwarfs}
\shortauthors{Filippazzo et al.}

\begin{document}


\title{Fundamental Parameters and Spectral Energy Distributions of Young and Field Age Objects with Masses Spanning the Stellar to Planetary Regime}


\author{{Joseph C. Filippazzo\altaffilmark{1,2,3}, Emily L. Rice\altaffilmark{1,2,3}, Jacqueline Faherty\altaffilmark{2,5,6}}, Kelle L. Cruz\altaffilmark{2,3,4}, Mollie M. Van Gordon\altaffilmark{7}, Dagny L. Looper\altaffilmark{8}}

\affil{\altaffilmark{1}Department of Engineering Science and Physics, College of Staten Island, City University of New York, 2800 Victory Blvd, Staten Island, NY 10314, USA}
\affil{\altaffilmark{2}Department of Astrophysics, American Museum of Natural History, New York, NY 10024, USA}
\affil{\altaffilmark{3}The Graduate Center, City University of New York, New York, NY 10016, USA}
\affil{\altaffilmark{4}Department of Physics and Astronomy, Hunter College, City University of New York, New York, NY 10065, USA}
\affil{\altaffilmark{5}Department of Terrestrial Magnetism, Carnegie Institution of Washington, DC 20015, USA}
\affil{\altaffilmark{6}Hubble Fellow}
\affil{\altaffilmark{7}Department of Geography, University of California, Berkeley, CA 94720, USA}
\affil{\altaffilmark{8}Tisch School of the Arts, New York University, New York, NY 10003, USA }




\begin{abstract}
We combine optical, near-infrared and mid-infrared spectra and photometry to construct expanded spectral energy distributions (SEDs) for 145 field age (\textgreater 500 Myr) and 53 young (lower age estimate \textless 500 Myr) ultracool dwarfs (M6-T9). This range of spectral types includes very low mass stars, brown dwarfs, and planetary mass objects, providing fundamental parameters across both the hydrogen and deuterium burning minimum masses for the largest sample assembled to date. A subsample of 29 objects have well constrained ages as probable members of a nearby young moving group (NYMG). We use 182 parallaxes and 16 kinematic distances to determine precise bolometric luminosities ($L_\text{bol}$) and radius estimates from evolutionary models give semi-empirical effective temperatures ($T_\text{eff}$) for the full range of young and field age late-M, L and T dwarfs. We construct age-sensitive relationships of luminosity, temperature and absolute magnitude as functions of spectral type and absolute magnitude to disentangle the effects of degenerate physical parameters such as $T_\text{eff}$, surface gravity, and clouds on spectral morphology. We report bolometric corrections in $J$ for both field age and young objects and find differences of up to a magnitude for late-L dwarfs. Our correction in $Ks$ shows a larger dispersion but not necessarily a different relationship for young and field age sequences. We also characterize the NIR-MIR reddening of low gravity L dwarfs and identify a systematically cooler $T_\text{eff}$ of up to 300K from field age objects of the same spectral type and 400K cooler from field age objects of the same $M_H$ magnitude.
\end{abstract}

\bibliographystyle{/usr/local/texlive/texmf-local/bibtex/bst/apj}





\section{Introduction}\label{sec:introduction}
Brown dwarfs are unable to sustain nuclear fusion in their cores due to insufficient mass and are thus degenerate across effective temperature, mass, and age. While these objects all contract to about the size of Jupiter within 500 Myr \citep{Bara98}, the extended photospheres of younger objects introduce the radius as yet another elusive observable. Entanglement of these fundamental parameters prohibits precise atmospheric characterization by spectral type and color alone. This necessitates determination of broader physical quantities such as distance, radius, and luminosity. Flux calibrated spectral energy distributions (SEDs) comprised of spectra as well as photometry enable precise empirical determination of $L_\text{bol}$ which can then be used to estimate additional stellar parameters.

Effective temperatures lower than about 3000K cause the emergent spectra of brown dwarfs to deviate substantially from that of a blackbody due to absorption and scattering from molecules, dust and clouds. Determination of their physical properties is further complicated as these substellar objects age and cool, changing opacity sources and evolving through later spectral types. Though current ultracool dwarf model atmosphere codes \citep{Alla13,Saum08,Burr11,Barm08} account for more complex chemistry and dynamics than ever before, incomplete physics and line lists frequently limit reliable data fits to those brown dwarfs which exhibit the simplest atmospheric conditions. Even for these objects, there are broad regions of the model spectra that have yet to reproduce observations and so must be excluded from the fitting routine \citep[e.g.][]{Cush08,Mann15}. Consequently, derivation of fundamental parameters with model atmospheres depend heavily on the included wavelength ranges, the resolution of the spectrum, the fitting technique, and the models used. 

\citet{Rice10a} fit model atmospheres to young ($\textless 10 Myr$) and field age late-M dwarfs and concluded that a combination of medium- and high-resolution NIR spectra is needed to derive fundamental parameters, though this is rarely attempted due to limited available data especially for fainter sources. \citet{Pati12} fit five model atmosphere grids to NIR spectra of a small sample of 1-50 Myr M8-L5 dwarfs and found $T_\text{eff}$ discrepant by up to 300K based on the models used. Despite such documented inconsistencies, this remains the most common way to extract atmospheric properties such as effective temperature, surface gravity, sedimentation efficiency, and metallicity \citep{Cush08,Step09,Witt11,Bonn14b,Diet14,Manj14}. Recent $L_\text{bol}$ determinations of large samples of M, L and T dwarfs \citep{Step09,Dupu13,Diet14,Schm14} have used MIR spectroscopy or photometry but have similarly employed various model atmospheres and fitting routines to estimate flux in areas without spectral coverage. Most other $L_\text{bol}$ estimates in the literature use bolometric corrections derived from these samples. These techniques are self-consistent in the parameters they predict but suffer from both known and unidentified systematics introduced by imperfect model atmosphere codes. Until the model grids reproduce the variety of our observations and fitting routines become more robust, a strictly empirical sanity check is needed. 

Direct integration of flux calibrated SEDs provides $L_\text{bol}$ as a function of more ubiquitous measurements such as magnitude, color, spectral index, and spectral type. Accurate characterization of these distance-scaled relationships can be powerful tools for inferring the atmospheric properties of additional ultracool dwarfs. Since distance is the dominant source of uncertainty in these calculations, the accumulation of trigonometric parallaxes for a diverse sample of late-M, L and T dwarfs \citep[e.g.][]{Dahn02,Tinn03,Vrba04,Fahe09,Dupu12a,Maro13,Diet14,Tinn14} is crucial to providing precise $L_\text{bol}$ measurements across the entire stellar/brown dwarf/planetary mass sequence. Radius and mass can then be inferred from from evolutionary models and $T_\text{eff}$ can be calculated from the Stefan-Boltzmann Law. This is preferable to deriving $L_\text{bol}$ from $T_\text{eff}$ values obtained by model atmosphere fits since the results are not tied to the fidelity of a fitting routine, the complexities of modeled atmospheric conditions, or the quality of the data.

Construction of ultracool dwarf SEDs from nearly complete observational coverage is therefore ideal and timely due to the mid-infrared (MIR) photometry of the Wide-Field Infrared Survey Explorer \citep[WISE;][]{Wrig10} and the \textit{Spitzer Space Telescope} Infrared Array Camera \citep[IRAC;][]{Fazi04}, as well as the groundswell of publicly available optical and infrared spectra of recent years \citep[e.g.][]{Cruz03,Cruz09,Burg14}. Calculation of $L_\text{bol}$ from direct integration of distance-scaled SEDs without the use of model atmosphere grids has been done in the past by \citet{Tinn93}, \citet{Legg01}, \citet{Dahn02}, \citet{Goli04a}, and \citet{Cush05} however these samples only had spectral and photometric coverage up to $4.1\mu$m.

We use $0.3-14.5\mu m$ spectra and photometry, and parallaxes and kinematic distances to create the largest collection of nearly complete flux calibrated ultracool dwarf SEDs assembled to date. In Sections \ref{sec:sample} and \ref{sec:data} we describe our sample and observational data used in the analysis. In Section \ref{sec:SEDs} we detail our SED construction procedure. In Section \ref{sec:Lbol} we describe our calculation of $L_\text{bol}$ for the sample and in Sections \ref{sec:radii}, \ref{sec:masses}, and \ref{sec:Teff} we determine radii, masses, and $T_\text{eff}$ respectively. In Section \ref{sec:discussion} we calculate bolometric corrections, construct luminosity-magnitude diagrams, and derive new spectral type-$T_\text{eff}$ relations in order to investigate the effects of age on the bulk physical properties of ultracool dwarfs. Conclusions are presented in Section \ref{sec:conclusions}.

\section{The Sample}\label{sec:sample}
Our goal was to assemble nearly complete SEDs to investigate global trends in the fundamental parameters of substellar objects, which presumably form like stars from the collapse of a shocked cloud of cold gas and dust. To accomplish this we constructed a sample of diskless objects no longer associated with a molecular cloud with masses just above the hydrogen burning minimum mass of about $79M_\text{Jup}$ all the way down to just below the deuterium burning minimum mass of about $12M_\text{Jup}$ \citep{Burr97,Chab00}. All objects were required to be within $80 pc$ to avoid interstellar extinction effects \citep{Aume09}.

Young objects are distinguished by their very red $J-K_S$ color due to dust and clouds in or above the photosphere \citep{Kirk06,Loop08b}, though not all very red objects have youth indicators \citep{Fahe13}. Therefore young dwarfs were identified based on probable membership in a NYMG \citep[Faherty et al. in prep; Reidel et al. in prep;][]{Gagn15a}. Additional young L0-L5 dwarfs were identified as those with a $\beta$ or $\gamma$ spectral type suffix \citep[][Cruz et al. in prep]{Kirk05,Kirk06,Cruz09,Rice10b} indicating low surface gravity features in the optical such as weak alkali doublets and weak metal hydride absorption. Late-M and late-L dwarf spectral types have been updated with $\beta/\gamma$ suffixes to reflect intermediate and very low surface gravity identified in the NIR \citep{Alle13}. In total, the sample contains 29 NYMG members and 24 low gravity dwarfs that were not placed in a NYMG.

We grouped all objects into three subsamples: 1) the core sample of 28 objects with a parallax measurement, optical through MIR photometry, and optical through MIR spectra, 2) the extended sample of 154 objects with the base requirements of a parallax, NIR spectrum, NIR photometry, and MIR photometry, and 3) the kinematic sample of 16 young objects with the same photometric and spectral requirements as the extended sample but with distances constrained by kinematics based on NYMG membership (Faherty et al., in prep). Optical photometry, optical spectra and/or MIR spectra were included for objects in the extended and kinematic samples where available. 

Our resulting sample is comprised of 198 ultracool dwarfs with spectral types of M6-T9 listed in Table \ref{table:sample_table} where we use optical spectral types for M and L dwarfs and IR spectral types for T dwarfs. We excluded Y dwarfs from the analysis due to the still emerging atmospheric complexities and poorly defined benchmarks for objects with $T_\text{eff}\textless 500K$. Fundamental properties of these latest type dwarfs will be addressed in Filippazzo et al., in preparation.

\section{The Data}\label{sec:data}
\subsection{Distances}\label{sec:distances}
Trigonometric parallaxes from the Brown Dwarf Kinematics Project \citep{Fahe12}, the Hawai'i Infrared Parallax Program \citep{Dupu12a}, and the literature \citep{Dahn02,Mone03,Tinn03,Dahn02,Vrba04,Duco08,Shko09,Kirk11,Liu_13,Maro13,Wein12,Zapa14a} with uncertainties better than 25$\%$ were used to calculate distances to the core and extended samples. For objects with multiple parallaxes, the value with the highest precision was used. The distance uncertainties calculated from these 182 parallaxes ranged from 0.3-25$\%$ with a mean of about 4$\%$. Parallaxes for the sample discrepant by more than $1\sigma$ are shown in Table \ref{table:parallaxes}.

\begin{deluxetable}{ccccc}
\tablecaption{Discrepant Parallaxes \label{table:parallaxes}}
\tablehead{Name                     &$\pi$                     &Ref       &Alt $\pi$          &Ref}\startdata
2M0501-0010             &51 $\pm$ 3.7             &1       &76.4 $\pm$ 4.8                      &2 \\
2M0355+1133             & 109.6 $\pm$ 1.3          &3       &133.7 $\pm$ 11.9                      &4 \\
LP 944-20 & 155.89 $\pm$ 1.03 & 5 & 201.4 $\pm$ 4.2 & 6
\enddata
\tablerefs{(1) \citet{Zapa14a}; (2) \citet{Fahe12}; (3) \citet{Liu_13a}; (4) \citet{Fahe13}; (5) \citet{Diet14}; (6) \citet{Tinn96}.}
\end{deluxetable}

Kinematic distances for 15 M and L dwarfs were calculated using the coordinates, proper motion, and radial velocity of the object given probable membership in a NYMG (Riedel et al., in prep; Faherty et al., in prep). A kinematic distance for GU Psc b was obtained from \citet{Naud14}. Uncertainties in these 16 kinematic distances were about 13$\%$.

\subsection{Photometry}\label{sec:photometry}
The primary sources of optical, NIR and MIR photometry were the Sloan Digital Sky Survey \citep[SDSS; ][]{York00}, the Two Micron All Sky Survey \citep[2MASS; ][]{Skru06}, and WISE respectively. Additional magnitudes came from the Deep Near-Infrared Survey of the Southern Sky \citep[DENIS; ][]{Epch97} and the literature using the Mauna Kea Observatory Near-Infrared \citep[MKO-NIR; ][]{Simo02,Toku02}, \textit{Spitzer Space Telescope} IRAC and Johnson-Cousins V(RI)$_C$ filter sets.

A total of 2360 magnitudes were available for 198 sources. Tables \ref{table:OPT_table}, \ref{table:NIR_table}, and \ref{table:MIR_table} show the optical, NIR, and MIR apparent magnitudes for the sample and the effective wavelength for each filter. We made no uncertainty cut on the data and uncertainties ranged from 0.06-13$\%$ with a mean of about 0.9$\%$.

\subsection{Published and Publicly Available Spectra}\label{sec:published_spectra}

We used 120, 193, and 53 previously published optical, NIR and MIR spectra, respectively, to construct SEDs for our sample. Table \ref{table:spectra_summary} shows a summary of the telescopes and instruments of origin along with the wavelength range, resolving power, and number of spectra used in our analysis for each. NIR spectra for about 77$\%$ of the sample are low resolution SpeX Prism \citep{Rayn03} data from the NASA Infrared Telescope Facility (IRTF), about half of which were obtained from the SpeX Prism Library$\footnote{http://www.browndwarfs.org/spexprism maintained by Adam Burgasser}$. 

The Low Resolution Imaging Spectrometer \citep[LRIS; ][]{Oke_95} instrument on Keck I was the source for 36$\%$ of all optical spectra used. All MIR spectra were obtained with the InfraRed Spectrograph \citep[IRS; ][]{Houc04} onboard the \textit{Spitzer Space Telescope} using the Short wavelength Low resolution (SL) module, 11 of which had Long wavelength Low resolution (LL) orders stitched in as well. Of the previously published IRS spectra, 26 are from the IRS Dim Suns project \citep{Roel04,Cush06,Main07} and 26 are from the IRS Enhanced Products Catalog of the Spitzer Heritage Archive  $\footnote{http://sha.ipac.caltech.edu/}$ (PIDs 2, 29, 51 and 30540 with PI J. Houck; PID 3136 with PI K. Cruz; PID 20409 with PI M. Cushing; PID 50367 with PIs M. Cushing and M. Liu; and PID 50059 with PI A. Burgasser).

\begin{deluxetable*}{cccccc}
\tablecaption{Instrument Summary of Spectra Used to Construct SEDs \label{table:spectra_summary}}\tablehead{\colhead{Instrument} & \colhead{Telescope} & \colhead{$\lambda (\mu m)$} & \colhead{R} & \colhead{Number} & \colhead{Refs}}
\startdata
\vspace{-0.2cm}\\
\multicolumn{6}{c}{Optical}\vspace{0.2cm}\\\tableline
LRIS & Keck I & 0.32$-$1.0 & 300$-$5000 & 43 & 1 \\
R$-$C Spec & KPNO 4m & 0.6$-$1.0 & 300$-$5000 & 29 & \nodata \\
R$-$C Spec & CTIO 4m & 0.55$-$1.0 & 300$-$3000 & 22 & \nodata \\
GoldCam & KPNO 2.1m & 0.55$-$0.93 & 300$-$4500 & 8 & \nodata \\
R$-$C Spec & CTIO 1.5m & 0.6$-$0.86 & 300$-$3000 & 4 & \nodata \\
FORS & ESO VLT U2 & 0.33$-$1.1 & 260$-$2600 & 3 & 2 \\
MagE & Magellan II Clay & 0.32$-$1.0 & $\sim$4100 & 3 & 3 \\
FOCAS & Subaru & 0.37$-$1.0 & 250$-$2000 & 2 & 4 \\
LDSS3 & Magellan II Clay & 0.6$-$1.0 & $\sim$1000 & 1 & 5 \\
DIS & ARC 3.5m & 0.38$-$1.0 & $\sim$600 & 1 & \nodata \\
\tableline
\vspace{-0.1cm}\\
\multicolumn{6}{c}{NIR}\vspace{0.2cm}\\\tableline
SpeX $-$ Prism & IRTF & 0.8$-$2.5 & $\sim$150 & 153 & 6 \\
SpeX $-$ SXD & IRTF & 0.8$-$2.4 & $\sim$2000 & 25 & 6 \\
SpeX $-$ LXD1.9 & IRTF & 1.95$-$4.2 & $\sim$2500 & 10 & 6 \\
GMOS$-$S & Gemini South & 0.36$-$0.94 & 670$-$4400 & 7 & 7 \\
FIRE $-$ Echelle & Magellan I Baade & 0.82$-$2.51 & $\sim$6000 & 7 & 8 \\
SpeX $-$ LXD2.3 & IRTF & 2.25$-$5.5 & $\sim$2500 & 2 & 6 \\
GMOS$-$N & Gemini North & 0.36$-$0.94 & 670$-$4400 & 2 & 7 \\
GNIRS $-$ SXD & Gemini North & 0.9$-$2.5 & $\sim$1700 & 2 & 9 \\
Triplespec & Palomar 200$-$inch & 0.95$-$2.46 & 2500$-$2700 & 2 & 10 \\
CGS4 & UKIRT & 4.5$-$5.0 & $\sim$400 & 2 & 11 \\
OSIRIS & CTIO 1.5m & 0.9$-$2.4 & 1400$-$3500 & 1 & 12 \\
SINFONI & ESO VLT U4 & 1.1$-$2.45 & 1500$-$4000 & 1 & 13,14 \\
NIRC & Keck I & 0.9$-$2.5 & $\sim$100 & 1 & 15 \\
IRCS & Gemini North & 1.2$-$2.4 & $\sim$100 & 1 & 16 \\
NIRI & Subaru & 1.0$-$2.5 & $\sim$500 & 1 & 17 \\
STIS & HST & 3.0$-$4.15 & $\sim$500 & 1 & 18 \\
\tableline
\vspace{-0.1cm}\\
\multicolumn{6}{c}{MIR}\vspace{0.2cm}\\\tableline
IRS $-$ SL & Spitzer & 5.1$-$14.5 & 60$-$128 & 51 & 19 \\
IRS $-$ LL & Spitzer & 14.5$-$37 & 60$-$128 & 11 & 19
\enddata
\tablerefs{(1) \citet{Oke_95}; (2) \citet{Appe98}; (3) \citet{Mars08}; (4) \citet{Kash00}; (5) \citet{Alli94}; (6) \citet{Rayn03}; (7) \citet{Alli02}; (8) \citet{Simc08}; (9) \citet{Elia06}; (10) \citet{Wils04}; (11) \citet{Moun90}; (12) \citet{DePo93}; (13) \citet{Eise03}; (14) \citet{Bonn04}; (15) \citet{Matt94}; (16) \citet{Koba00}; (17) \citet{Hoda03}; (18) \citet{Wood98}; (19) \citet{Houc04}.}
\end{deluxetable*}

More than half of the objects had multiple optical and NIR spectra so we selected one from each regime with the broadest wavelength coverage and highest signal to noise (S/N). These spectra were trimmed of exceptionally noisy edges by truncating the ends up to the first wavelength position with S/N$\ge$20. To generate an uncertainty array for the 15 optical spectra where one was not available, we used S/N=5 for all wavelengths. Even this very conservative estimate had no effect on the resulting fundamental parameters since the uncertainty in the distance dominates after the spectra are flux calibrated. 

\subsection{New NIR and MIR Spectra}\label{sec:new_opt_nir_spectra}

We present new low-resolution SpeX Prism data for 2MASSI J1017075+130839, 2MASS J23224684-3133231, and 2MASS J05012406-0010452 taken on UT 2011 December 08, 2006 August 28, and 2007 October 12, respectively. We took 180s exposures at two nod positions along the slit, which was oriented to the parallactic angle to minimize slit loss and spectral slope variations. The raw images were corrected for non-linearity, pair subtracted, and flat fielded. We obtained spectra of A0 dwarf stars for telluric correction and flux calibration.  A set of exposures of internal flat field and argon arc lamps were also taken for flat fielding and wavelength calibration. Data were then reduced with the SpeXtool package \citep{Cush04} using standard settings. 

We also present new medium-resolution NIR spectra taken with the TripleSpec \citep{Wils04} instrument on the 200-inch Hale Telescope at Palomar Observatory. LSPM J1658+7027 and 2MASSW J1155395-372735 were observed on 2010 June 03 and 2009 January 28, respectively. We used an ABBA nod sequence with exposures times of about 300s and observed nearby A0 dwarf stars to perform telluric correction and flux calibration. Dome flats were taken to calibrate the pixel-to-pixel response and data were reduced using a version of Spextool modified for Triplespec.

And finally, we present a new MIR spectrum for 2MASS J21392676+0220226 obtained on 2005 May 25 (PID 3136, PI K. Cruz) with the IRS instrument using the SL module in standard staring mode. Target observations consisted of eight 60 second exposures. The raw data was then put through the Basic Calibrated Data (BCD) processing pipeline, which flags and masks hot pixels, performs a droop correction, subtracts a dark sky reference frame, and does a flat-field correction to account for detector irregularities and spectral order discrepancies. The spectrum was extracted from the background subtracted images with SSC{'}s SPICE program. SPICE extracts a 1D spectrum from the input image using the determined position of the trace and the wavelength-dependent Point Spread-Function (PSF) (SPICE Spitzer IRS Custom Extraction v. 1.4). Because the standard extraction window width resulted in false features and other artifacts in the spectrum, we used a custom extraction window width of 4 pixels at 7$\mu m$ which worked well with our data. After extraction with SPICE, custom IDL scripts were used to co-add the exposures and merge the separate orders.

\section{Construction of SEDs}\label{sec:SEDs}
\subsection{Creation of Composite Optical and NIR Spectra}\label{sec:composite_spectra}
Overlapping optical and NIR spectra were available for 120 of the 198 sample objects. We combined these into a single composite spectrum using the shared red optical wavelength range to maximize the broadband filter coverage for flux calibration. We smoothed the higher resolution spectrum to the lower resolution wavelength array in the area of overlap and normalized the two components. Since the spectra have different methods of telluric correction or were not telluric corrected at all, the $0.93-0.97\mu m$ range was excluded from the normalization process to avoid the telluric feature in that region. The composite spectrum was defined as the mean flux value at each resampled point weighted by the inverse of the uncertainty. A new uncertainty array was created for the composite spectrum by summing component uncertainties in quadrature. \citet{Cush05} find that no significant errors are introduced due to the difference in relative flux densities when creating composite spectra as described.

Given the luxury of multiple optical and NIR resolutions for many sources, we tested whether the integrated flux calculated from the optical+NIR composite spectrum was sensitive to our choice of input spectra. For 52 sources we constructed composites with every permutation of available optical and NIR spectra for a total of 304 tests. We integrated under the curve along the longest common wavelength baseline to get the total flux in the area of overlap of each source's set of SEDs. The total flux for any source in the optical and NIR is effectively resolution independent in the range $75\textless R\textless 5000$, varying at most by 0.4$\%$. Table \ref{table:spectra_table} shows the final list of spectra used to construct each SED in the sample.

\subsection{Flux Calibration of Spectra}\label{sec:flux_calibration}
As described in Section \ref{sec:data}, spectra used in our SEDs were taken on different days with different instruments under different seeing conditions so there is little uniformity of the data. To account for this heterogeneity we flux calibrate all spectra to available SDSS, V(RI)$_C$, 2MASS, DENIS, IRAC, and WISE photometry.  We shifted the apparent magnitudes to absolute magnitudes using parallaxes or kinematic distances and converted to photon flux densities in units $erg s^{-1} cm^{-2} \mbox{\AA}^{-1}$ using

\begin{equation}\label{eqn:mag-to-flux} 
f_\lambda=f_{ZP}\cdot 10^{-\frac{M}{2.5}}
\end{equation}

where $M$ is the absolute magnitude and $f_{ZP}$ is the flux of a Kurucz model of Vega$\footnote{http://kurucz.harvard.edu/stars.html}$ at the filter effective wavelength listed in Tables \ref{table:OPT_table}, \ref{table:NIR_table}, and \ref{table:MIR_table}. The SDSS magnitude system \citep{Fuku96} is not quite in the AB magnitude system \citep{Oke_83} so we converted first to AB using the corrections of \citet{Holb06} and then to the Vega system using the corrections of \citet{Blan07}. Uncertainties introduced from these corrections were negligible.

To calibrate the spectra to the absolute photometry, we calculated synthetic magnitudes for every band with complete spectrum coverage using 

\begin{equation}\label{eqn:syn-mag}
m=-2.5\log\left(\frac{\int f_\lambda S(\lambda)\lambda d\lambda}{\int f_{Vega} S(\lambda)\lambda d\lambda}\right)
\end{equation}

Here $f_\lambda$ and $f_{Vega}$ are the photon flux densities of the spectrum and Vega respectively and $S(\lambda)$ is the filter response function.

\begin{figure}[t]
\begin{center}
\plotone{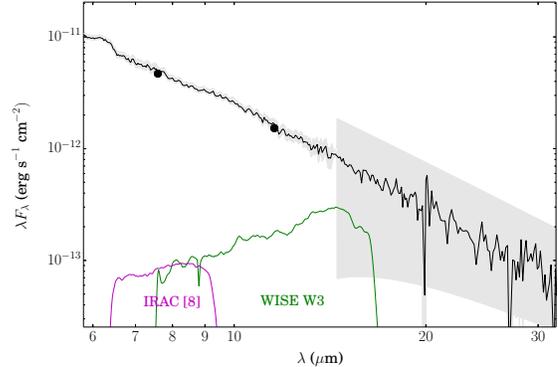}
\caption{\label{fig:IRS-calibration}The $5.2-37\mu m$ IRS spectrum of Gliese 752 B (black line) shows that appending a Rayleigh-Jeans tail with a 300K uncertainty in the blackbody temperature (grey shaded region) accounts for all variation in the spectrum long ward of $14.5\mu m$. While all IRS spectra cover the IRAC [8] passband (magenta curve), we use this approximation to calculate synthetic W3 magnitudes for IRS spectra that do not cover the entire W3 passband (green curve).}
\end{center}
\end{figure}

IRS spectra were converted from $F_\nu$ in $Jy$ to $F_\lambda$ in $erg s^{-1} cm^{-2} \mbox{\AA}^{-1}$ via the relationship $F_\nu \nu=F_\lambda \lambda$. Most IRS spectra in our sample cover the wavelength range $5.25-14.5\mu m$ so we used \textit{Spitzer} IRAC channel 4 centered at $8\mu m$ (hereafter [8]) and/or WISE W3 centered at $12\mu m$ to maximize the number of points available for flux calibration. If a source had a magnitude in W3 but the IRS spectrum ended at $14.5\mu m$, we appended a Rayleigh-Jeans tail from the longest wavelength value out to the end of the W3 filter at $17.26\mu m$. The uncertainty on the tail was set as a 300K uncertainty in the blackbody curve.  Though the flux in the range $14.5-17.26 \mu m$ constitutes 28$\%$ of the synthetic magnitude in W3, the relative lack of features at $\lambda\textgreater 14.5\mu m$ \citep{Cush06} and our conservative estimate of uncertainty makes this a good approximation. The 11 IRS spectra for our sample with coverage out to $37\mu m$ (spectral types M6.5, M8, M8.5, M9, L0.5, L1, L3, L3.5, L5, and T7) confirm this estimate. Figure \ref{fig:IRS-calibration} shows the [8] and W3 filter response curves along with a typical IRS SL/LL spectrum and a Rayleigh-Jeans tail appended at $14.5\mu m$ to cover the entire W3 band for comparison. The W3 synthetic magnitude was then calculated using equation \ref{eqn:syn-mag}.

For each non-contiguous piece of the spectrum, we calculated a normalization constant C by minimizing the ratio of the catalog and synthetic photometry as

\begin{equation}\label{eqn:calibration}
C=\frac{\sum\limits_i f_i\hspace{5pt}f_{i\text{, syn }}\Big/\left(\sigma^2_i+\sigma^2_{i\text{, syn}}\right)}{\sum\limits_i f_i^2\Big/\left(\sigma^2_i+\sigma^2_{i\text{, syn}}\right)},
\end{equation}

where $f_i$ and $f_{i\text{, syn}}$ are the catalog and synthetic fluxes respectively for the i-th photometric band, and $\sigma_i$ and $\sigma_{i\text{, syn}}$ are their respective uncertainties. Each spectrum segment was then multiplied by its respective constant C to anchor it to the maximum wavelength baseline of absolute catalog photometry rather than just a single photometric band.

\subsection{Filling in the Gaps}\label{sec:gaps}
To calculate $L_\text{bol}$ from a complete SED, we filled in the gaps between the non-contiguous flux calibrated spectra and survey photometry, linearly interpolated to zero from the short wavelength limit of our data, and appended a Rayleigh-Jeans tail to the long wavelength limit of our data. The flux in regions with photometry but no spectra were estimated by linearly interpolating between magnitudes. The 1.4, 1.9 and 2.8 $\mu m$ telluric $H_2O$ absorption features were also linearly interpolated across at the flux levels on either side of the gap if necessary. 

SDSS ugriz photometry was available for 73 objects, which allowed us to linearly interpolate down to $0.35\mu m$ before linearly interpolating to zero flux at zero wavelength on the Wein tail of the SED. $V(RI)_C$ photometry for 37 objects was also included. For all other objects, we estimated ugriz photometry based on absolute SDSS-2MASS magnitude-magnitude relations derived from the sample. We fit 3rd-order polynomials to $M_z$ versus $M_J$, $M_H$, and $M_{Ks}$ and used the relationship with the smallest dispersion to infer an $M_z$ magnitude. The rms of the polynomial fit was used as the uncertainty on the estimated absolute magnitude. This process was repeated for the other four SDSS bands. $M_u$, $M_g$, $M_r$, $M_i$, and $M_z$ were estimated for each object based on their value of $M_{Ks}$, $M_J$, $M_H$, $M_H$, and $M_J$ respectively. 

IRAC [3.6], [4.5], [5.8], and [8] photometry was available for 72 objects that also had WISE W1, W2, and W3 photometry, which we used to derive IRAC-WISE magnitude-magnitude relations to better estimate the flux in the MIR. Using the relationship with the smallest rms we estimated $M_{[3.6]}$, $M_{[4.5]}$, $M_{[5.8]}$, and $M_{[8]}$ for each object based on their value of $M_{W1}$, $M_{W2}$, $M_{W1}$, and $M_{W1}$ respectively. For the 7 objects with no W1 photometry, we estimated $M_{W1}$ using $M_{[3.6]}$, $M_{L'}$, or $M_{Ks}$ based upon availability of those magnitudes in that order. $M_{W2}$ was similarly estimated for those seven objects using $M_{[4.5]}$, $M_{L'}$, or $M_{Ks}$. We estimated $M_{W3}$ for 24 objects in the same way using $M_{[8]}$ or $M_{W2}$. $M_{L'}$ was estimated using $M_{[3.6]}$ or $M_{W1}$. Since it is well documented that the colors of ultracool dwarfs are sensitive to age \citep{Kirk06,Kirk08,Cruz09,Fahe09,Schm10,Alle10,Biha10,Fahe12}, we derived one polynomial for the field age objects ($M_\text{FLD}$) and a separate one for objects with signatures of low surface gravity, probable membership in a NYMG, or youth indicators of a stellar companion ($M_\text{YNG}$). The exceptions were u-, g-, and L'-band, which had either too few magnitudes or too large a scatter about the field sequence to justify distinct FLD and YNG polynomials. All magnitude-magnitude relations mentioned above are presented in Table \ref{table:polynomials}. 

Longward of the longest wavelength MIR data point, we appended a Rayleigh-Jeans tail out to $1000 \mu m$. If the SED had an IRS spectrum, the tail was flux calibrated to the wavelength range $11\textless\lambda\textless 14.5 \mu m$. This was done to avoid the $CH_4$ ($9.2\textless\lambda\textless 14.5 \mu m$) and $NH_3$ ($10.25\textless\lambda\textless 11 \mu m$) absorption features in late-L and T dwarfs \citep{Marl96,Burr01,Roel04} and the $9\textless\lambda\textless 11 \mu m$ plateau caused by small silicate grains in the upper cloud decks of some L dwarfs \citep{Cush06}. If no IRS spectrum was available, the Rayleigh-Jeans tail was flux calibrated to the W3 magnitude, which largely avoids these features as well. The uncertainty on the Rayleigh-Jeans tail was conservatively set as a 300K uncertainty in the blackbody temperature, accounting for all variation past W3 band in the 11 SEDs with spectral coverage out to $37\mu m$.

\begin{figure}[t]
\begin{center}
\plotone{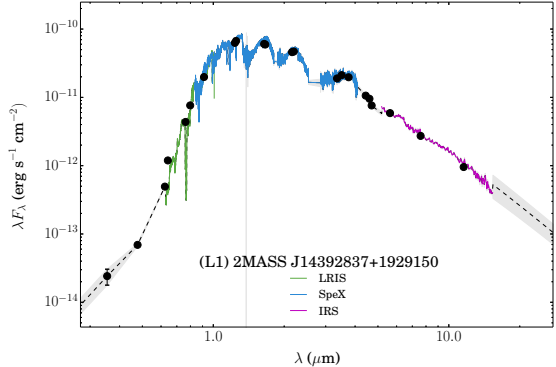}
\caption{\label{fig:core-SED}Core sample flux calibrated SED of 2MASS J14392837+1929150 showing the optical (green), NIR (blue) and MIR (magenta) spectra used. Black points are photometry, the dashed black line shows how gaps were filled in regions with no spectra, and the grey shaded region is the uncertainty.%
}
\end{center}
\end{figure}

A complete SED is shown in Figure \ref{fig:core-SED} where optical, NIR and MIR spectra are shown in blue, green and magenta respectively. Photometric points are displayed as black markers, the curve integrated under to calculate $L_\text{bol}$ is the black dashed line, and the grey region is the $1\sigma$ uncertainty of the SED.

\section{Bolometric Luminosities}\label{sec:Lbol}
We calculated $L_\text{bol}$ for all the sources in our sample by integrating under the absolute flux calibrated SED from $0-1000\mu m$. We used

\begin{equation}\label{eqn:Lbol}
L_\text{bol}=4\pi d^2 \int\limits_{0\mu m}^{1000\mu m} F_\lambda d\lambda,
\end{equation}

where $F_\lambda$ is the absolutely flux calibrated SED and $d$ is the distance in parsecs to the source. All values of $L_\text{bol}$ in this work are listed in units of $\log (L_\text{bol}/ L_\odot )$ where we use $L_\odot =3.846\times 10^{33}\text{ erg } \text{s}^{-1}$ \citep{Cox_00}. 

\begin{figure*}[t!]
\begin{center}
\plotone{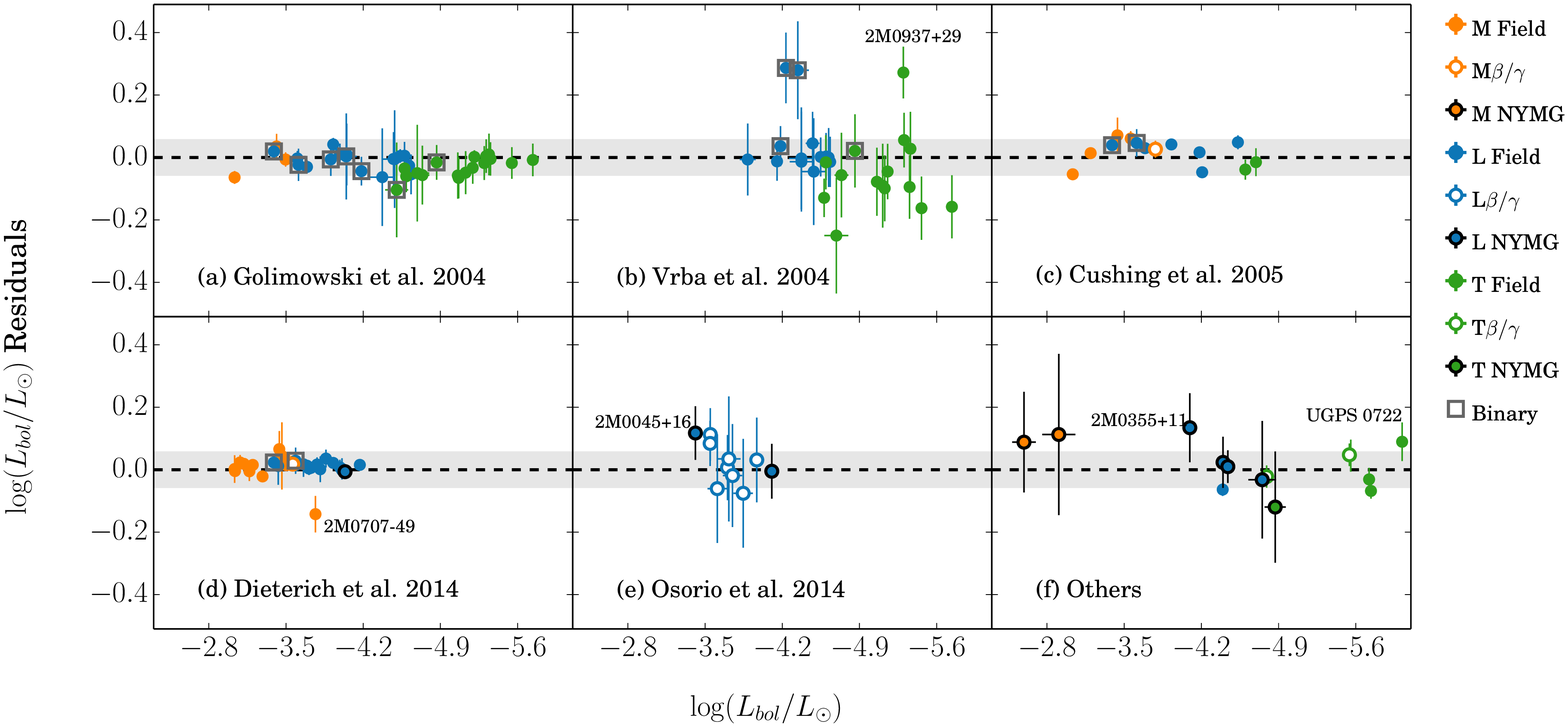}
\caption{\label{fig:Lbol-comparison}Residuals of $L_\text{bol}$ calculated in this work compared to the values presented in (a) \citet{Goli04a}, (b) \citet{Vrba04}, (c) \citet{Cush05}, (d) \citet{Diet14}, (e) \citet{Zapa14a}, and (f) smaller samples from the literature. M, L and T dwarfs are shown as orange, blue, and green points, respectively. Field age objects are filled points with no outline, members of a NYMG are filled points with a black border, and objects with signatures of youth or low surface gravity $\beta/\gamma$ designations that could not be placed in a NYMG are unfilled points. Binaries are indicated by a grey square.%
}
\end{center}
\end{figure*}

We present the first bolometric luminosities for 65 field age and 21 low gravity objects. Table \ref{table:Lbol_comparison_table} shows the remaining 112 objects with each previously published value of $L_\text{bol}$, the parallax used in that work, and the $L_\text{bol}$ value we obtained using our method with the same parallax. Panels a-f in Figure \ref{fig:Lbol-comparison} show our values compared with those of \citet{Goli04a}, \citet{Vrba04}, \citet{Cush05}, \citet{Diet14}, \citet{Zapa14a}, and other smaller samples respectively.

Our values of $L_\text{bol}$ agree with those of \citet{Cush05} (Figure \ref{fig:Lbol-comparison}c) to within $1\sigma$, which is to be expected since our method of calculation is very similar to the one presented there. However, Cushing et al. append a Rayleigh-Jenas tail longward of L' band at $4.1\mu m$ since the sample predates the MIR capabilities of WISE and Spitzer. While our values of $L_\text{bol}$ agree very well, our analysis of the precision attainable without MIR data suggests that their uncertainties are slightly underestimated (See Section \ref{sec:spectral_coverage}). Similarly, \citet{Goli04a} assumed a Rayleigh-Jeans tail longward of L'- or M'-band for their M and L dwarfs which produces good agreement with our early- to mid-L dwarf luminosities. They use the $0.6-4.1\mu m$ SED of the T6.5 dwarf Gl 229B to estimate a correction for their late-T dwarfs and use half that correction for their early-T dwarfs. While this correction results in similar $L_\text{bol}$ values as our nearly complete late-T SEDs, halving the correction for the early-T dwarfs overestimates the flux longward of $2.5\mu m$ causing all the bright T dwarfs in Figure \ref{fig:Lbol-comparison}a to fall within the uncertainties but below our distribution.

\citet{Vrba04} use the K-band bolometric correction derived by \citet{Goli04a} to estimate $L_\text{bol}$ for their sample, causing a larger dispersion and a similar overbrightness in the T dwarfs shown in Figure \ref{fig:Lbol-comparison}b. We find the slightly blue 2MASS J09373487+2931409 (T6) about 30\% brighter and three other T-dwarfs about 20\% dimmer. While this bolometric correction in Ks may prove useful for M and L dwarfs, these objects demonstrate the danger of a catch-all correction of late type ultracool dwarfs given their large dispersion in color for objects of the same spectral type. 

\citet{Diet14} fit synthetic magnitudes of BT-Settl 2010 model atmospheres to $VRI+JHK+W1W2W3$ photometry to derive $L_\text{bol}$ for their sample of mid-M to mid-L dwarfs. The good agreement with our empirical results show this method to be fairly robust in this range, though the exclusion of late-L and T dwarfs in the Dieterich sample limit its utility to early type brown dwarfs. Though in agreement with the results of this work, most objects systematically lie just above (slightly dimmer than) our findings. \citet{Diet14} acknowledge that their results are about 100K cooler than the effective temperatures derived by \citet{Goli04a} using direct integration and evolutionary models, so perhaps this temperature disagreement is due to their method underestimating the total flux. The source of the 12$\%$ discrepancy for the outlier 2MASS J07075327-4900503 (M9) is unclear but likely the result of a poor atmospheric model fit to the data in that work.

\citet{Zapa14a} have a sample of 10 suspected young L dwarfs for which they determine $L_\text{bol}$. They use $BC_{Ks}=3.40$ of \citet{Todo10}, derived from three young M9.5-L0 dwarfs, for their four L0-L2 dwarfs. For the six L3-L5 dwarfs in their sample, they use $BC_{Ks}=3.22$ and $BC_J=1.16$ determined for G196{-}3B (L3$\beta$) in \citet{Zapa10}. The disagreement of the three brighter objects in Figure \ref{fig:Lbol-comparison}e is due to their application of a larger $BC_{Ks}$ for young objects than for field age objects, while we find that low surface gravity results in a larger dispersion and not necessarily an offset from the field age correction (See Section \ref{sec:BCs} for discussion on BCs for young ultracool dwarfs).

Figure \ref{fig:Lbol-comparison}f shows $L_\text{bol}$ for a variety of objects in the literature using different methods of calculation. \citet{Legg12} use $0.6-4.2\mu m$ spectral coverage and a bolometric correction derived from model atmospheres for UGPS J072227.51-054031.2, which we find about 10$\%$ brighter. For the remarkably red AB Dor member 2MASS J03552337+1133437, \citet{Liu_13} use direct integration of a flux calibrated NIR spectrum coupled with a model atmosphere fit to place this object about 12\% dimmer than our result. Just as with the \citet{Vrba04} and \citet{Zapa14a} samples, the use of bolometric corrections for many of these smaller samples generally greatly increases the uncertainty and scatter of the predicted $L_\text{bol}$ values.

Distance is by far the largest source of uncertainty in these measurements making large, high-precision parallax programs an imperative for revealing the bulk properties of ultracool dwarfs. Additionally, the large scatter in color due to secondary characteristics such as metallicity, dust, and clouds make values of $L_\text{bol}$ unreliable when derived using spectrophotometric distances or bolometric corrections insensitive to this diversity. We find that direct integration of patchy flux calibrated SEDs produce the smallest uncertainties in $L_\text{bol}$ of as little as 2$\%$ (See Section \ref{sec:spectral_coverage}).

\section{Radii}\label{sec:radii}
Though long-baseline interferometry has been used to measure radii for stars as late as M4 \citep{Boya12}, ultracool dwarfs are as yet too small and dim for this technique. Lacking a significant number of substellar radii measurements, atmospheric and evolutionary models are commonly used instead. One method is to fit a grid of model atmospheres to spectra or photometry and then scale the emitted flux of the best fitting model to the absolute flux calibrated spectrum.  A radius can then be estimated from the observed flux if the distance to the object is known via the $(R/d)^2$ scale factor. However, incomplete line lists and poorly reproduced regions such as the H-band peak and $\lambda\textless 0.9 \mu m$ result in sometimes highly discrepant best fit parameters (e.g. $T_\text{eff}$ and surface gravity) and consequently unreliable or unphysical radii estimates \citep[e.g.][]{Bowl10,Barm11,Dupu13,Liu_13}. We derive parameters for our sample using this technique in Filippazzo et al. (in preparation), for comparison with the semi-empirical results of this paper. 

The method we employ in this work is to use $L_\text{bol}$ and the object age to determine the radius from evolutionary models. We assume an age of 0.5-10 Gyr for all field objects in the sample that have no signatures of youth. Since electron degeneracy pressure limits the size of all field age ultracool dwarfs to about that of Jupiter, the dispersion in radius even over this colossal timespan is predicted to be about $0.1R_\text{Jup}$. Objects younger than this are likely still undergoing gravitational contraction and are not yet completely electron degenerate, as evidenced by low surface gravity spectral features compared to field age objects. While objects optically typed $\beta /\gamma$ exhibit signatures of low surface gravity including weaker absorption and pressure broadening of alkali lines \citep{Cruz03}, ages cannot yet be accurately inferred solely from a gravity suffix \citep[][; Faherty et al. in prep]{Cruz09,Alle13,Zapa14a}. 

Ages of nearby stellar associations have been constrained using a variety of techniques such as dynamical age, Li abundance, $H\alpha$ emission, X-ray emission, and color-magnitude diagrams. Therefore, we have identified 29 objects in our sample that are likely members of NYMGs and adopted the age of the parent association as the age of the object. We use the age ranges and references therein presented by \citet{Malo12} in their Table 1 of 8-20 Myr for TW Hydrae (TWA), 12-22 Myr for $\beta$ Pictoris ($\beta$Pic), 30-50 Myr for Argus, 50-130 Myr for AB Doradus (AB Dor), and 10-40 Myr for Tucana-Horlogium (Tuc-Hor), Columba, and Carina.

Ages for GL 337CD, Gl 417BC, HN Peg B, HD 3651B, Ross 458C, Luhman 16AB, 2MASS J22344161+404138, and G 196-3B were obtained from the literature using X-ray luminosity, kinematics, chromospheric activity, Li abundance, and/or gyrochronology of a stellar companion \citep{Kirk01b,Burg05,Luhm07a,Gold10,Fahe14,Shko09}. For the 17 $\beta /\gamma$ objects in our sample that do not have age estimates and could not be placed in NYMGs but are too low-g to be field age, we used an age range of 8-130 Myr to reflect the minimum and maximum ages of the considered NYMGs. 

\begin{figure}[t]
\begin{center}
\plotone{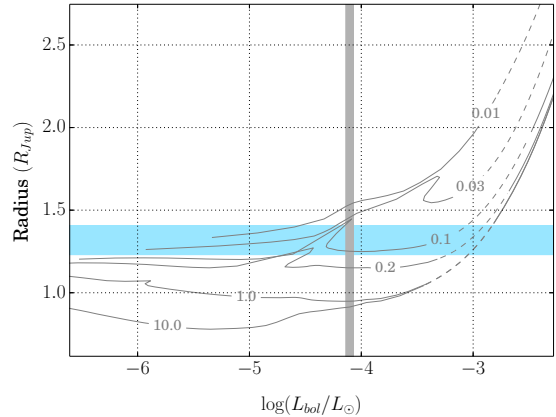}
\caption{\label{fig:radius-interp}Example of how the radius of 2MASS 0355+1133 was chosen by interpolation between evolutionary model isochrones based on the age of its parent association AB Doradus. The solid grey lines to the left and right of the dashed grey lines show solar metallicity SMHC08 and DMESTAR isochrones in Gyrs, respectively. The dashed grey lines show our interpolation used to make the isochrones continuous. The grey vertical bar shows the 1$\sigma$ $L_\text{bol}$ value and the horizontal blue bar shows the resulting radius range.%
}
\end{center}
\end{figure}

We used the inferred age and calculated $L_\text{bol}$ for each object to find the range of predicted radii from the solar metallicity, hybrid cloud (SMHC08) evolutionary model isochrones of \citet{Saum08} (Figure \ref{fig:radius-interp}). We used these model tracks since they they do not treat grain sedimentation efficiency ($f_{sed}$) as a free parameter exhibiting cloudy late-M through L/T transition dwarfs and cloud-free mid- to late-T dwarfs, in agreement with observations. 

Differences in gas and condensate chemistry, molecular opacities, cloud modeling, and atmospheric boundary conditions \citep{Saum08} make radii assumptions heavily tied to the models being used. Thus for M6-T3 dwarfs, we also computed the radii predicted by the evolutionary models of \citet{Chab00} (DUSTY00) and the $f_{sed}=2$ (SMf208) models of \citet{Saum08}. The final radius range for each source was set as the minimum and maximum values of all model predictions for the given $L_\text{bol}$ and age. The same approach was used for L6-T8 dwarfs using the models of \citet{Bara03} (COND03) and the cloudless (SMNC08) models of \citet{Saum08}. None of the models mentioned above extend to sufficiently high luminosity for us to extract radii for our brightest objects. In order to include the most luminous late-M dwarfs of our sample, we used the Dartmouth Magnetic Evolutionary Stellar Tracks and Relations \citep[DMEstar;][]{Feid12,Feid13} and performed a cubic interpolation across the gaps of each isochrone to make them continuous.

The few directly measured substellar radii have relied upon transit duration and orbital velocity measurements of double-lined eclipsing systems. Unfortunately, none of these objects meet the spectral and photometric requirements to be included in our sample, though we discuss them here for comparison with the predicted values of this work. Excluding M dwarfs and young objects since they are potentially not electron degenerate, we find a range of $0.78-1.14R_\text{Jup}$ for 95 L0-T9 dwarfs with assumed ages of $0.5-10Gyr$ and solar metallicity. The directly measured radii of the $0.4-12.7Gyr$ brown dwarfs OGLE-TR-122b \citep{Pont05}, CoRoT-3 b \citep{Dele08},  CoRoT-15b \citep{Bouc10}, WASP-30 b \citep{Ande10}, LHS 6343 C \citep{John11}, KELT-1 b \citep{Sive12}, KOI-205 b \citep{Diaz13}, KOI-415 b \citep{Mout13}, and SDSS J141126.20+200911.1 \citep{Litt14} produce a range of $0.66-1.19R_\text{Jup}$. Though in very good agreement, even these few directly measured radii exhibit a range almost 70\% larger than that predicted by model isochrones. This is most likely a result of some unaccounted for physical processes in the evolutionary models and/or non-solar metallicity of the objects resulting in a greater size diversity.

\section{Masses}\label{sec:masses}
\begin{figure*}[t!]
\begin{center}
\plotone{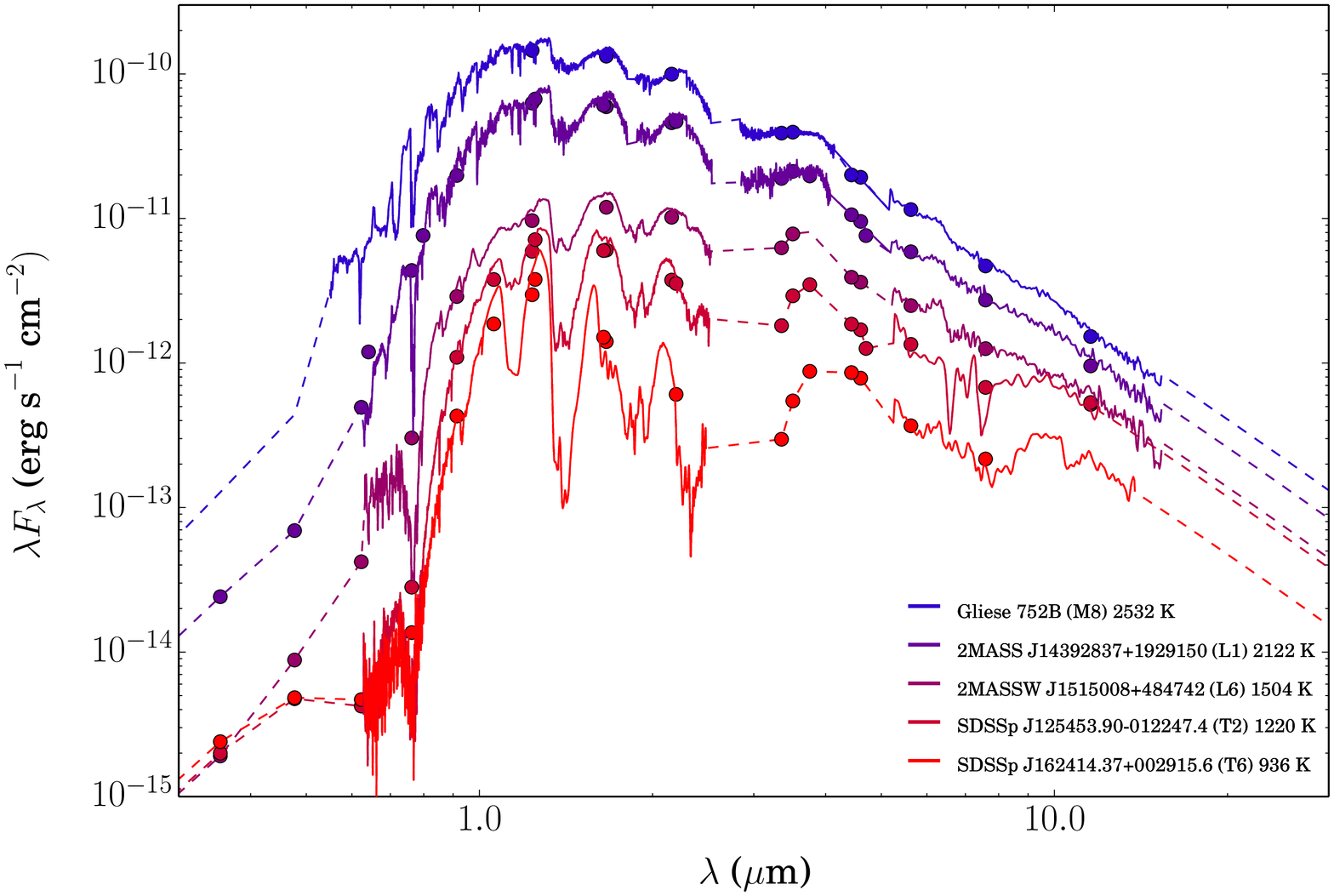}
\caption{\label{fig:field-sequence}Flux calibrated SEDs of field age dwarfs Gliese 752B (M8), 2MASS J14392837+1929150 (L1), 2MASSW J1515008+484742 (L6), SDSSp J125453.90-012247.4 (T2), and SDSSp J162414.37+002915.6 (T6). Solid lines are observed spectra, points are photometry, and dashed lines show how gaps in the data were filled. Uncertainties were removed for clarity. %
}
\vspace{1cm}
\plotone{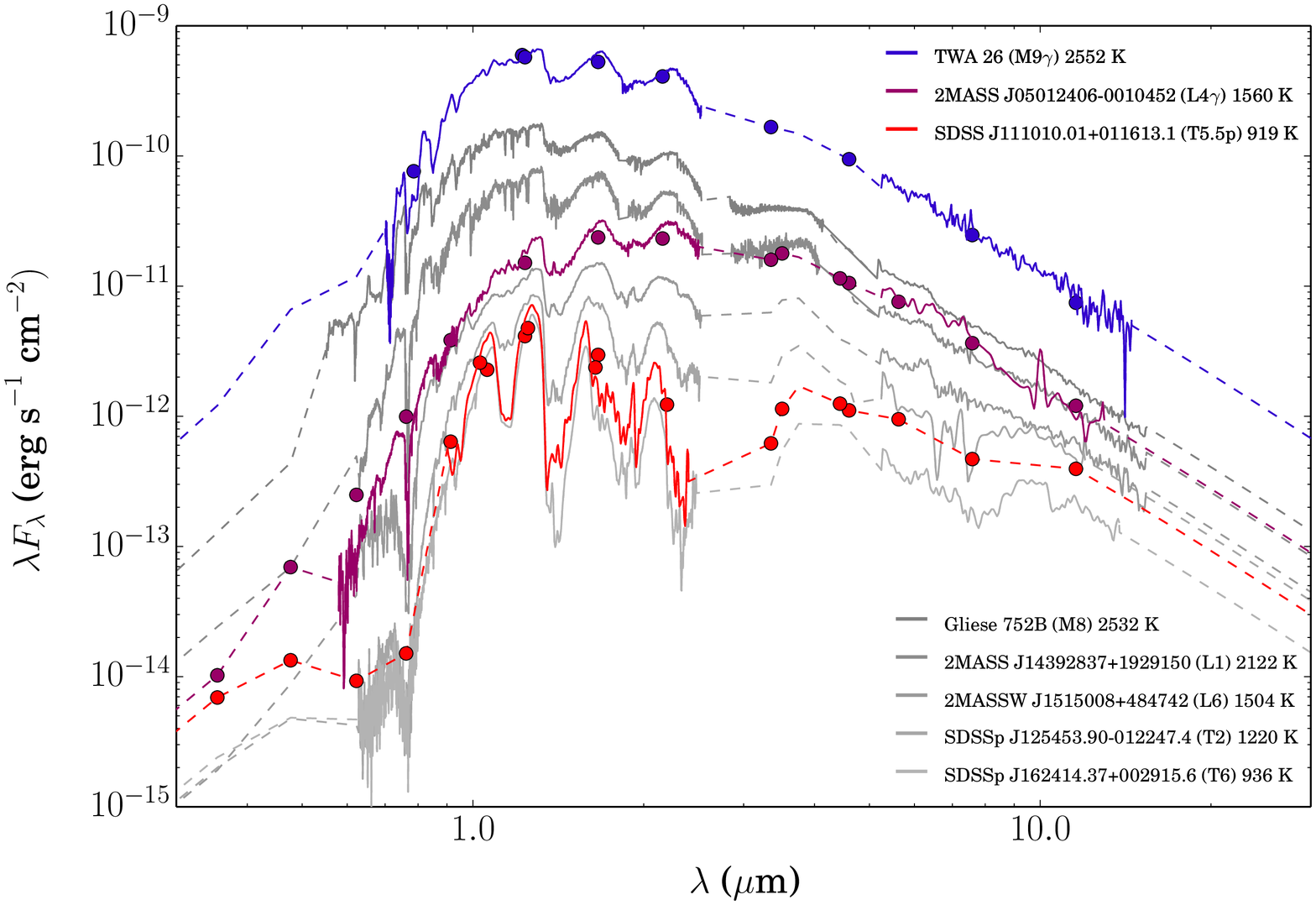}
\caption{\label{fig:young-sequence}Flux calibrated SEDs of TWA 26 (M9$\gamma$), 2MASS J05012406-001045 (L4$\gamma$), and SDSS J111010.01+011613.1 (T5.5p AB Dor member) plotted with the field age sequence shown in Figure \ref{fig:field-sequence}. Symbols are the same as in Figure \ref{fig:field-sequence}.%
}
\end{center}
\end{figure*}

\begin{figure*}[t!]
\begin{center}
\plotone{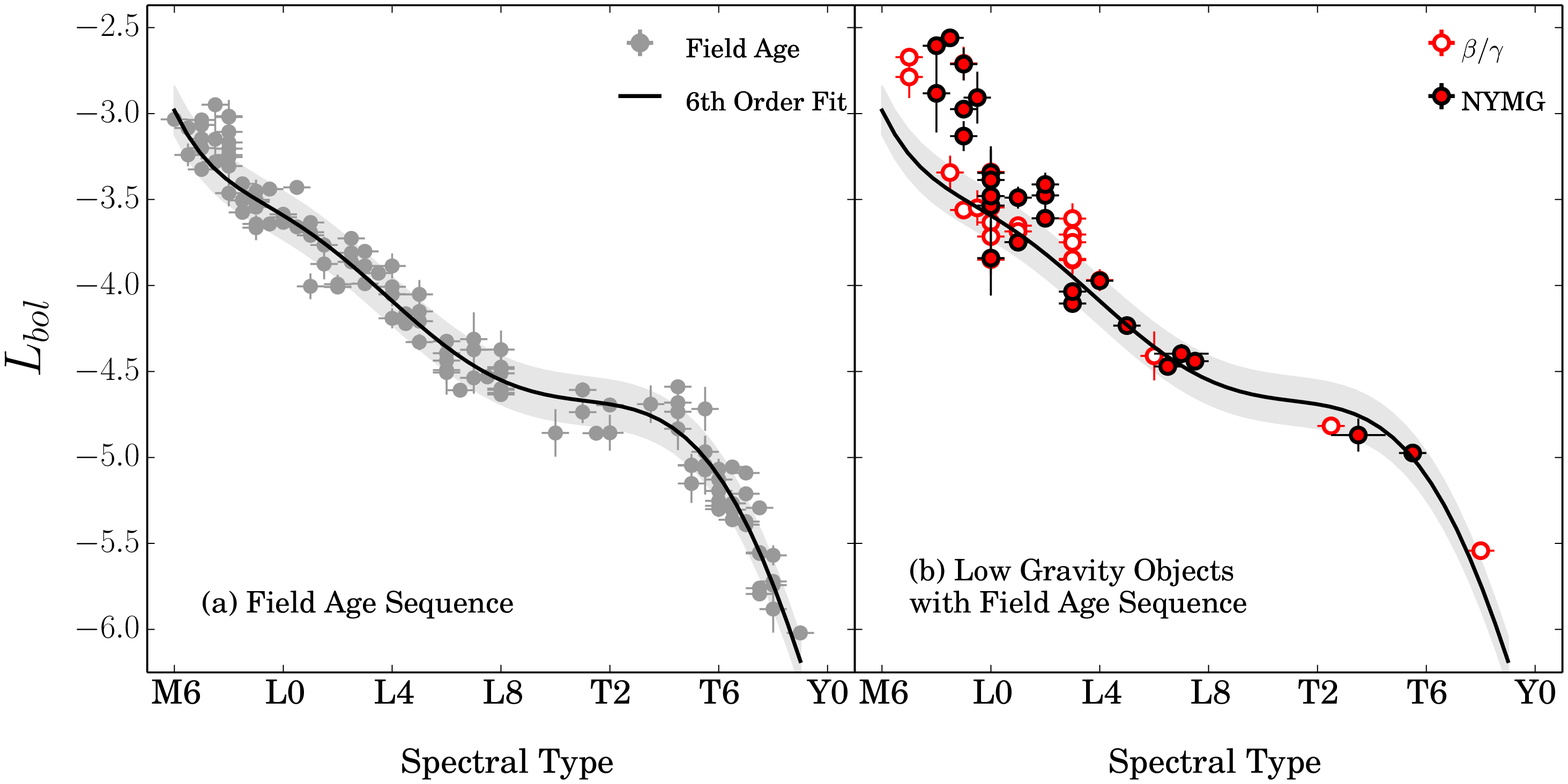}
\caption{\label{fig:Lbol-v-SpT}$L_\text{bol}$ as a function of spectral type for (a) 124 field age objects (grey points) and (b) 48 objects with probable membership in a NYMG (red points with a black border) or signatures of low surface gravity (unfilled points with a red border). A 6th order weighted polynomial fit (solid black line) to the field objects only is shown in both plots.%
}
\end{center}
\end{figure*}

The mass of an ultracool dwarf determines all its other fundamental parameters as a function of time yet it remains an unobservable for single objects. While dynamical masses of visual binaries point to possibly problematic cooling rates of substellar isochrones \citep{Dupu09b,Dupu09c,Dupu10,Kono10}, we are only equipped to infer model-derived masses given the data on hand.

We employ the same technique that was used to estimate radii from evolutionary model isochrones in Section \ref{sec:radii} to estimate masses for the sample as well. Just as with the radius, we use our calculated $L_\text{bol}$ and the maximum and minimum predicted masses of the SMHC08, SMf208, and DUSTY00 tracks for M6-T3 dwarfs and the SMHC08, SMNC08, and COND03 tracks for L6-T8 dwarfs to infer a mass range. We find that the results are highly dependent on the model isochrones and produce uncertainties in mass as low as 3$\%$ and as high as 75$\%$. Masses for the sample, which span both the hydrogen burning and deuterium burning limits, are given in Table \ref{table:params_table}.

\section{Effective Temperatures}\label{sec:Teff}
We calculated $T_\text{eff}$ for each object using our inferred radius $R$, integrated $L_\text{bol}$, and the Stefan-Boltzmann Law

\begin{equation}\label{eqn:Teff}
T_\text{eff}=\left(\frac{L_\text{bol}}{4\pi R^2\sigma_{SB}}\right)^\frac{1}{4},
\end{equation}

where $\sigma_{SB}$ is the Stefan-Boltzmann constant. We present fundamental parameters including $L_\text{bol}$, radius, mass, surface gravity, and $T_\text{eff}$ for the entire sample in Table \ref{table:params_table}.

Uncertainties in $T_\text{eff}$ are due primarily to the uncertainty in the object distance and radius estimates from evolutionary models. As \citet{Dupu13} point out, subtle differences in radii do not have a large effect on the calculated temperature since $T_\text{eff}\propto R^{-1/2}$. To a lesser degree, our assumptions about the shape and uncertainty of the Wien tail in the optical, linear interpolation between photometry, and flux calibration and uncertainty estimates of the Rayleigh-Jeans tail in the MIR also contributed. Typical uncertainty in $T_\text{eff}$ for our sample is about 6$\%$.

\section{Discussion}\label{sec:discussion}

\subsection{Flux Calibrated SED Sequences}
To demonstrate the broad changes in the SED with spectral type, we plot a sequence of flux calibrated field age object SEDs shown in Figure \ref{fig:field-sequence}. For comparison, Figure \ref{fig:young-sequence} shows the same sequence with three young flux calibrated SEDs over plotted. The M9$\gamma$ TWA 26 is redder in OPT-NIR color and 80\% brighter than the equal temperature M8 Gleise 752B. The L4$\gamma$ 2MASS J05012406-0010452 has as much flux in the optical as the equal temperature L6 2MASSW J1515008+484742. However, the young object is as bright in the MIR as the L1 2MASS J14392837+1929150, which is almost 600K hotter. The T5.5p AB Dor member SDSS J111010.01+011613.1 has a very similar SED at all wavelengths and is only 20\% brighter than the field age T6 SDSSp J162414.37+002915.6. If this trend holds, young T dwarfs would be almost indistinguishable from field age objects of the same spectral type given the dispersion in $L_{bol}$ of about 15\% (Section \ref{sec:relations}).

\subsection{$L_\text{bol}$-Magnitude-Spectral Type Relations}\label{sec:relations}
Figure \ref{fig:Lbol-v-SpT}a shows $L_\text{bol}$ vs. spectral type for 124 field age objects with our weighted 6th-order polynomial fit (Table \ref{table:polynomials}) and rms of $0.14$ dex displayed as a solid black line and grey shaded region respectively. Luminosity decreases almost linearly from M6-L7 with a typical scatter in a given spectral type of about 30$\%$. $L_\text{bol}$ is almost constant through the L/T transition as condensates rain out of the photosphere, quickly evolving through spectral type as deeper and brighter layers are exposed \citep{Marl10}. The luminosity then drops steeply through the late-T dwarfs. 

Figure \ref{fig:Lbol-v-SpT}b shows $L_\text{bol}$ vs. spectral type for 26 probable members of a NYMG and 22 objects with signatures of low surface gravity. The 6th-order polynomial fit to the field age objects is displayed again for comparison, which shows almost all young late-M and L dwarfs lie on or slightly above this sequence due to their still contracting radii. The almost order of magnitude dispersion in our empirically determined luminosities of young objects at the M/L transition demonstrates the predicted sharp decline in radius for young objects with $L_\text{bol}$ greater than $\sim -3.5$  shown in Figure \ref{fig:radius-interp}. In the most extreme case, the brightest young M9 in the sample (TWA 26) is over three times brighter than the dimmest (LP 944-20). The brightest young L0 (2MASS J01415823-4633574) is 60$\%$ brighter than the dimmest (2MASS J02411151-0326587). The three young T dwarfs HN Peg B, GU Psc B, and SDSS J111010.01+011613.1 lie on or below the field age sequence while most young L dwarfs lie on or above it. A larger sample of bona fide young T dwarfs is needed to explore this relationship at later types.

\begin{figure}[t]
\begin{center}
\plotone{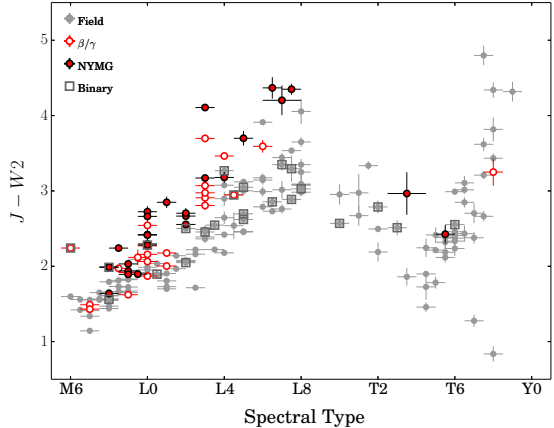}
\caption{\label{fig:J-W2}J-W2 color as a function of spectral type for 135 field, 22 low gravity, and 26 NYMG members. NYMG members and low gravity L dwarfs are up to a full magnitude redder than field age objects of the same spectral type. Symbols are the same as in Figure \ref{fig:Lbol-v-SpT}.%
}
\end{center}
\end{figure}

\begin{figure}[t]
\begin{center}
\plotone{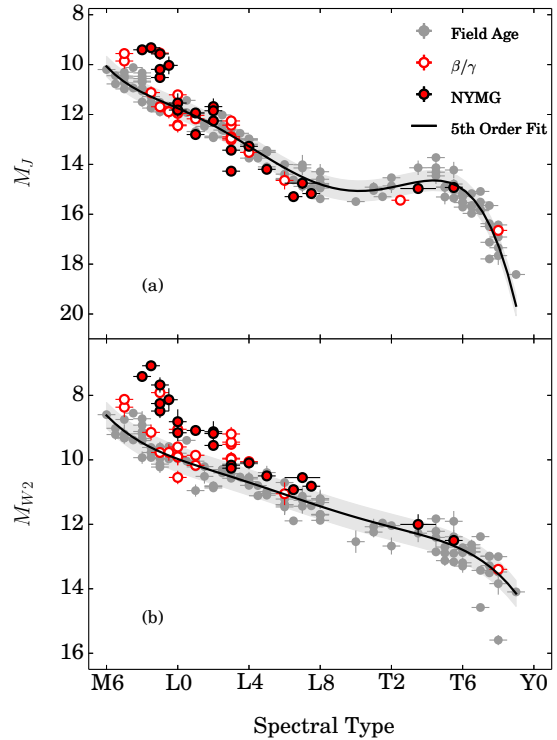}
\caption{\label{fig:spt-v-m}(a) $M_J$ as a function of spectral type shows 12 young late-M dwarfs are up to 2 magnitudes brighter than the 5th order weighted polynomial fit (black line) of 115 field age objects. The 30 young L and T dwarfs have very similar or slightly dimmer $M_J$ magnitudes compared to the field age sequence. (b) $M_W2$ as a function of spectral type shows 42 young M, L and T dwarfs to be about 1.5, 0.5 and 0.25 magnitudes brighter than the 4th order weighted polynomial fit of 117 field age objects. The rms of the polynomial fit is shown as the grey shaded region in both plots. Symbols are the same as in Figure \ref{fig:Lbol-v-SpT}.%
}
\end{center}
\end{figure}

Figure \ref{fig:J-W2} shows how young L dwarfs have J-W2 colors a full magnitude redder than their field age counterparts possibly due to scattering of light in their dusty extended photospheres \citep{Cruz09,Gizi12,Fahe12,Fahe13}. This is in agreement with \citet{Zapa14a} who report comparable $M_J$ magnitudes for 10 young L dwarfs with the field age sequence of \citet{Dupu12a}. \citet{Fahe12} find 7 young M dwarfs $\textgreater 0.5$ mag brighter and 10 young L dwarfs 0.2-1.0 mag dimmer in J-band, though the reddening discussed in that work is in NIR color. With MIR data for our subsample of 42 young objects, we can confirm if this effect is due to an under or over brightness in one of the bands or a true shifting of flux out to longer wavelengths. We find that young objects are 0.5-1.5 mags brighter in $M_{W2}$ than field age objects of the same spectral type (Figure \ref{fig:spt-v-m}a), while their $M_J$ magnitudes range from 2 mags brighter for late-M and 1 mag dimmer for late-L dwarfs, crossing the field age sequence at L0 (Figure \ref{fig:spt-v-m}b).

\begin{figure*}[t]
\begin{center}
\plotone{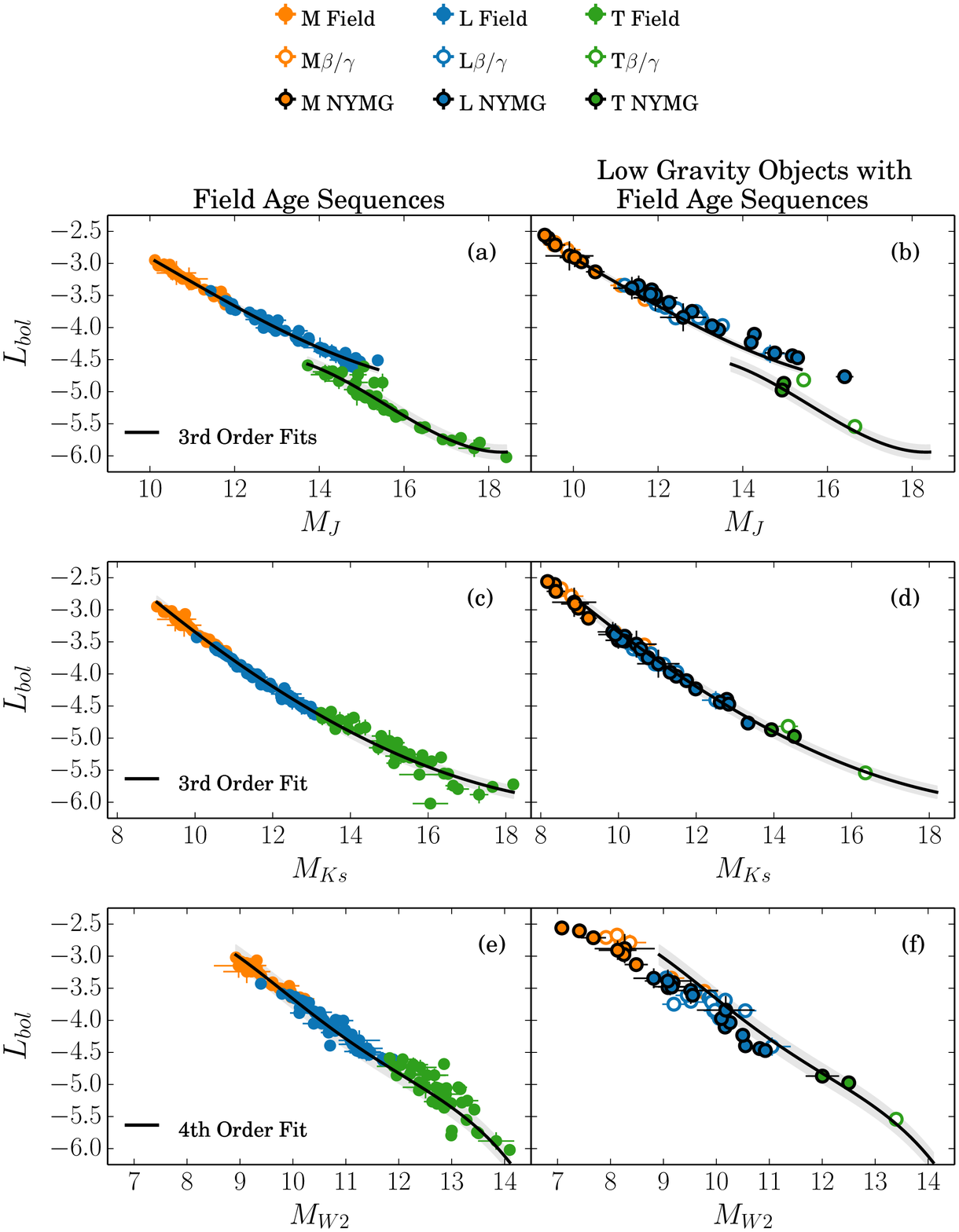}
\caption{\label{fig:M-v-Lbol}$L_{bol}$ as a function of $M_J$ (top row), $M_{Ks}$ (center row), and $M_{W2}$ (bottom row). Figures a, c, and e show field age objects with 3rd, 3rd, and 4th order polynomial fits, respectively. Figures b, d, and f show those same field sequences with young objects over plotted. Symbols are the same as in Figure \ref{fig:Lbol-comparison}.}
\end{center}
\end{figure*}

When compared in terms of $L_\text{bol}$, however, low gravity L dwarfs distinguish themselves from normal L dwarfs as systematically dimmer in $M_J$ (Figure \ref{fig:M-v-Lbol}a,b) and brighter in $M_{W2}$ (Figure \ref{fig:M-v-Lbol}e,f), corresponding to a redder NIR-MIR color while maintaining about the same bolometric luminosity \citep{Zapa14a}. 

\begin{figure}[h!]
\begin{center}
\plotone{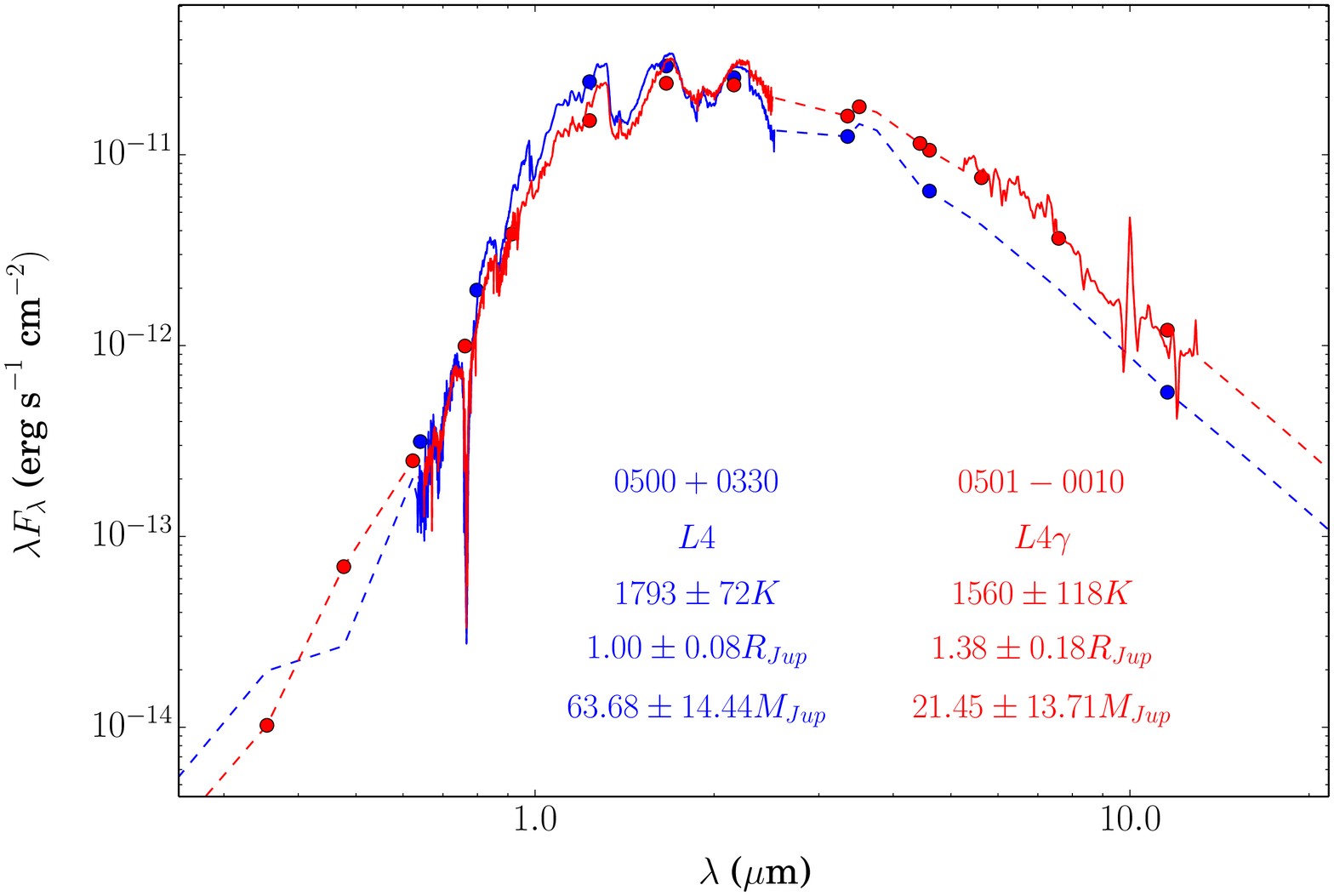}
\caption{\label{fig:young-old} Flux calibrated SEDs of 2MASS J05002100+0330501 (L4; blue) and 2MASS J05012406-0010452 (L4$\gamma$; red) demonstrate objects with equal luminosities but substantially different colors, radii, temperatures, and masses.%
}
\end{center}
\end{figure}

Figure \ref{fig:young-old} shows the field age L4 dwarf 2MASS J05002100+0330501 and the redder L4$\gamma$ 2MASS J05012406-0010452, which have nearly equal $L_\text{bol}$. The L4$\gamma$ however is 0.53 mags brighter in $M_{W2}$ and 0.51 mags dimmer in $M_J$. The SEDs of these objects suggest the flux pivots around H or Ks band, shifting from $\lambda\textgreater 2.5\mu m$ to $\lambda\textless 1.2\mu m$ as an L dwarf ages \citep{Fahe13}. This color mismatch for objects of the same luminosity implies that a bolometric correction (BC) in J or W2 is different for young and field age objects as suggested by \citet{Todo10}, \citet{Zapa10}, \citet{Luhm12}, \citet{Fahe13}, \citet{Liu_13a} and \citet{Zapa14a}.

\subsection{Bolometric Corrections}\label{sec:BCs}
Imprecise bolometric corrections are the primary cause of the large scatter of objects in Figures \ref{fig:Lbol-comparison}b, e, and f, which can propagate into highly inaccurate fundamental parameters for ultracool dwarfs. With strictly empirical $L_\text{bol}$ values for such a large sample and the routine use of these corrections in the literature, we revisited $BC_J$ and $BC_{Ks}$ for both young and field age objects using $M_{\text{bol}\odot}=+4.74$. 

\begin{figure}[t]
\begin{center}
\plotone{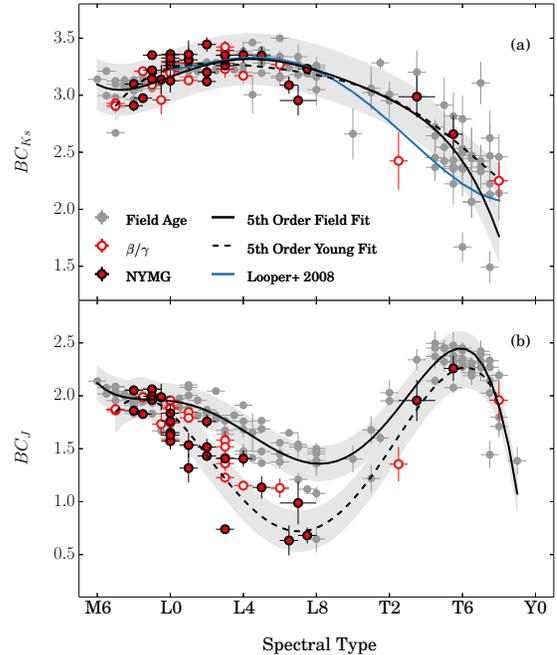}
\caption{\label{fig:BCs}Bolometric corrections for the sample are shown in $Ks$ (a) and $J$ (b) bands for low gravity (dashed line) and field age (solid line) objects as a function of spectral type. The \citet{Loop08b} relation is shown as the blue line. Symbols are the same as in Figure \ref{fig:Lbol-v-SpT}.%
}
\end{center}
\end{figure}

$M_{Ks}$ has an almost linear dependence on $L_\text{bol}$ for late-M and L dwarfs of all ages in the sample (Figure \ref{fig:M-v-Lbol}c,d). This makes $M_{Ks}$ an ideal band from which $L_\text{bol}$ can be determined for late-M and L dwarfs with no age information. We derived $BC_{Ks\text{ FLD}}$ by fitting a 5th order polynomial to 122 field age objects. Figure \ref{fig:BCs}a shows our relation in $Ks$ is in excellent agreement with that of \citet{Loop08b} \citep[][relation excluding binaries; magenta line]{Goli04a} for L0-L8 dwarfs but the dispersion of up to 1 mag in mid- to late-T dwarfs causes a mismatch of the polynomials at later types. This may be due to ours being a more complete sample containing twice as many T dwarfs as \citet{Loop08b} and/or their oversimplification of MIR flux for T dwarfs based solely on the L' and M' spectra of GL 229B.

\citet{Todo10} find that $BC_{Ks}$ for three young M9.5/L0 dwarfs is about 0.2 mags larger than that of field age L0 objects and claim that a different $BC_{Ks}$ is needed for young and old objects. While our results agree to within $1\sigma$ for the two common sample objects 2MASS J01415823-4633574 and 2MASS J02411151-0326587, we derive $BC_{Ks\text{ YNG}}$ and find corrections for 8 other young L0 dwarfs only 0.05 mags larger on average than field age objects and well within the uncertainties of the field sequence.

We also derive $BC_{J\text{ FLD}}$ and $BC_{J,\text{ YNG}}$ for the sample as a function of spectral type in Figure \ref{fig:BCs}b and Table \ref{table:polynomials}. The low gravity sequence is remarkably different than the field age sequence, differing by as much as a full magnitude for late-L dwarfs of the same spectral type. The tighter correlation of spectral type with $BC_J$ rather than $BC_{Ks}$ for mid- to late-T dwarfs suggests that the former is actually a more reliable correction to use when estimating late-type luminosities for field age objects. More confirmed young T dwarfs are needed to confirm $BC_{J\text{ YNG}}$ at later types.

\subsection{Dependence of $L_\text{bol}$ on Spectral Coverage}\label{sec:spectral_coverage}
Given the sample's varying degrees of spectral coverage, linear interpolation was used between photometric points where no spectra were available. To justify our estimations, we investigated the sensitivity of the apparent bolometric magnitude ($m_\text{bol}$) to the amount and quality of data available in order to create a prescription for determining reliable $L_\text{bol}$ values when certain data are missing.

\begin{figure}[t]
\begin{center}
\plotone{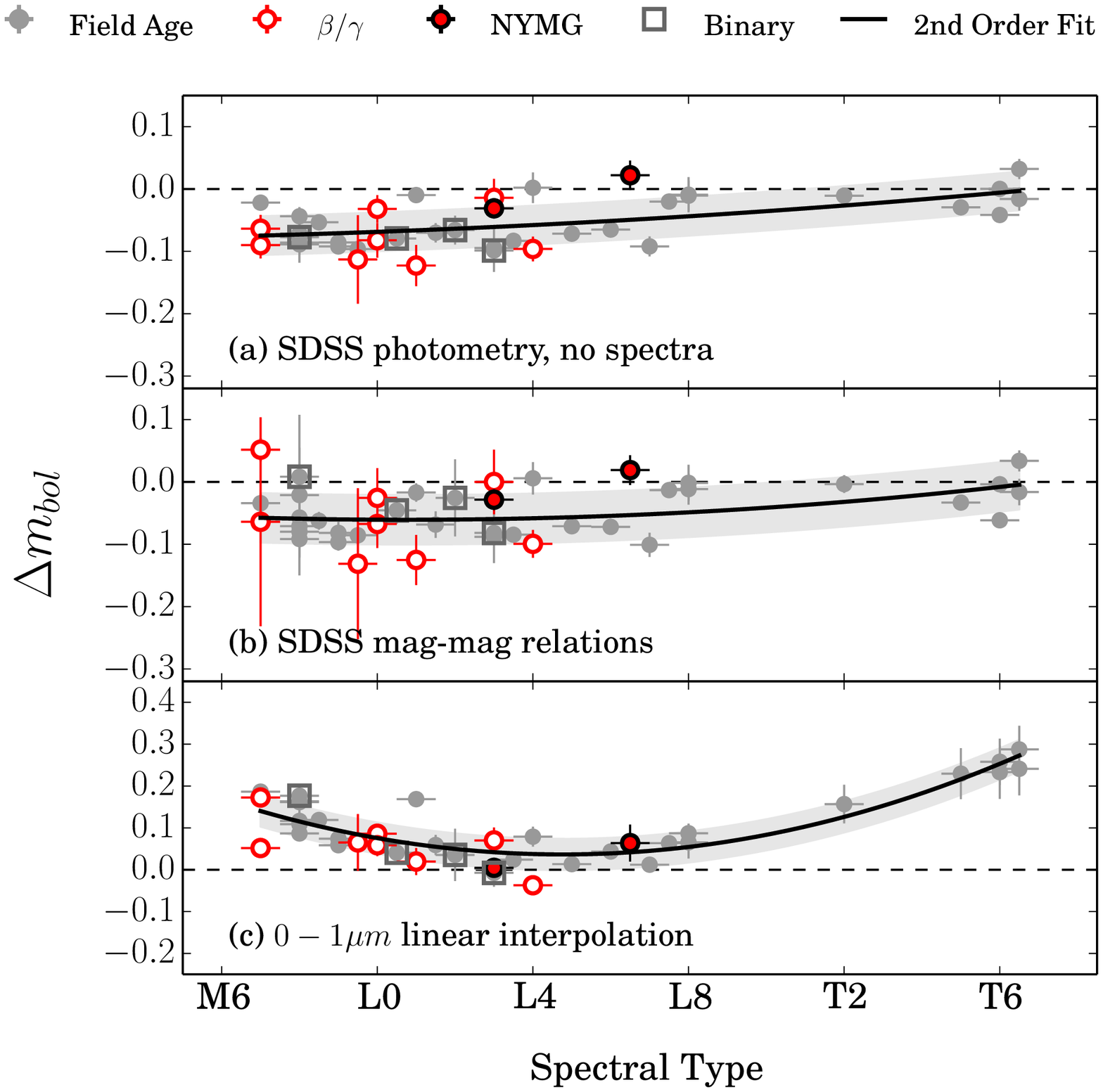}
\caption{\label{fig:opt_coverage}Dependence of $m_\text{bol}$ on different data coverage scenarios in the optical for 44 field age objects: (a) linear interpolation through SDSS photometry, (b) linear interpolation through photometry estimated from magnitude-magnitude relations, and (c) linear interpolation from $1\mu m$ down to zero flux at zero wavelength. Symbols are the same as in Figure \ref{fig:Lbol-v-SpT}.%
}
\vspace{0.5cm}
\plotone{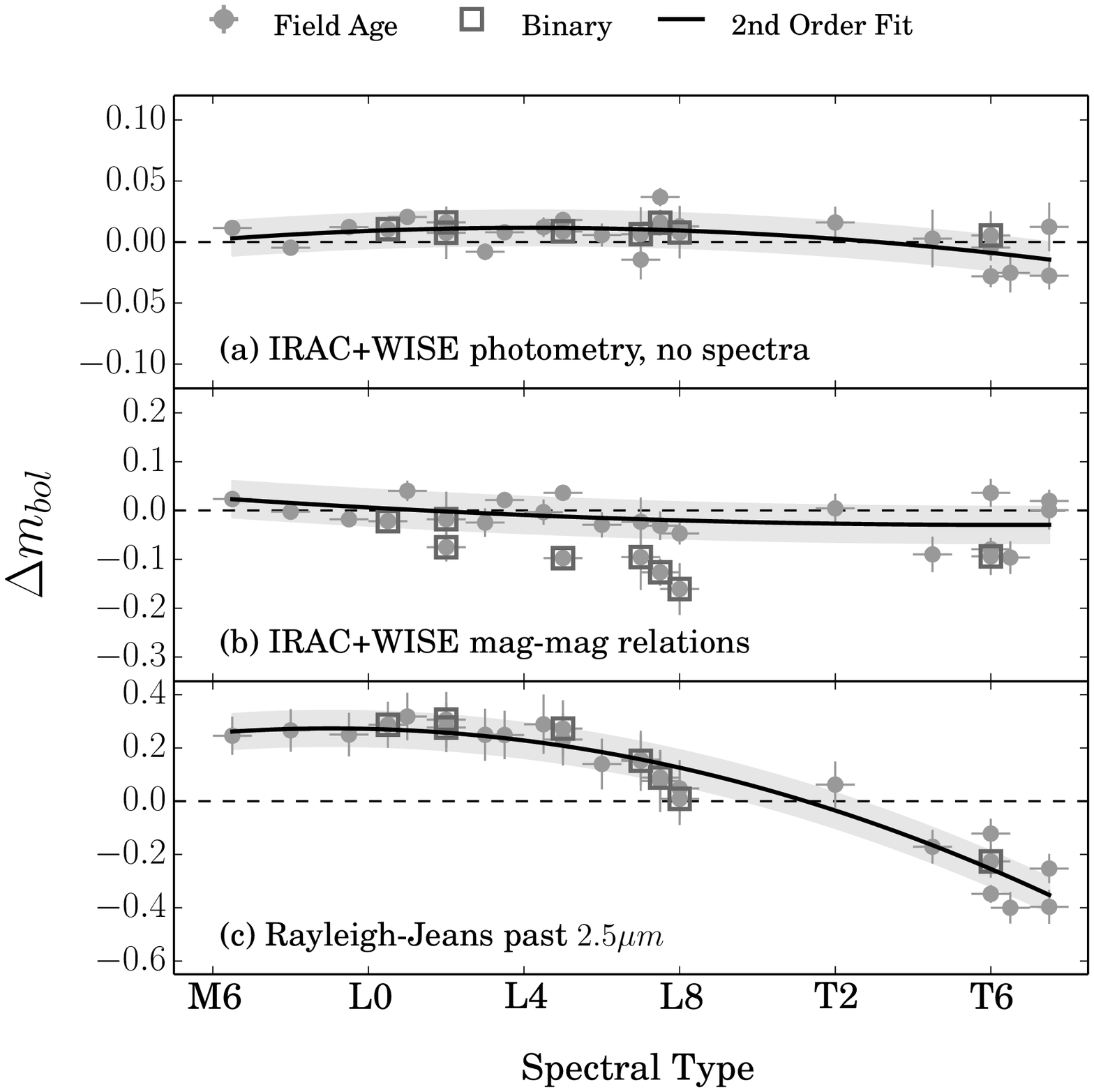}
\caption{\label{fig:mir_coverage}Dependence of $m_\text{bol}$ on different data coverage scenarios in the MIR for 30 field age objects: (a) linear interpolation through IRAC+WISE photometry, (b) linear interpolation through IRAC+WISE photometry estimated from magnitude-magnitude relations, and (c) appending a Rayleigh-Jeans tail long ward of $2.5\mu m$. Symbols are the same as in Figure \ref{fig:Lbol-v-SpT}.%
}
\end{center}
\end{figure}

To perform this test in the optical, we selected the 41 objects with $0.6\textless\lambda\textless 1\mu m$ spectral coverage and ugriz photometry and determined the change in total flux under three different optical data scenarios. Figure \ref{fig:opt_coverage}a shows the difference between $m_\text{bol}$ measured with optical spectra and $m_\text{bol}$ measured with linear interpolation through SDSS photometry as a function of spectral type. Even though the flux is overestimated when the deep $KI$ doublet and $VO$ bands in the $0.73\textless\lambda\textless 0.8\mu m$ region are washed out, the curve drawn by linear interpolation through SDSS photometry lies below the optical pseudo continuum. This causes a measured $m_\text{bol}$ about 0.1 mags lower for late-M and L dwarfs. This technique reproduces the total optical flux for T dwarfs very well with an uncertainty of only 0.05 mags.

In the absence of optical spectra and measured ugriz photometry, Figure \ref{fig:opt_coverage}b shows that 2MASS JHKs absolute magnitudes can be used to very accurately reproduce SDSS photometry using the magnitude-magnitude relations presented in Table \ref{table:polynomials}. The almost identical Figures \ref{fig:opt_coverage}a and \ref{fig:opt_coverage}b show that linear interpolation through estimated optical photometry performs just as well as flux calibrated ugriz magnitudes. 

Figure \ref{fig:opt_coverage}c shows the difference in $m_\text{bol}$ between an SED with an optical spectrum and one where linear interpolation was used from $1\mu m$ all the way down to zero flux at zero wavelength. This method overestimates the flux shortward of $1\mu m$ by up to 0.2 mags for late-M and L dwarfs. For late-T dwarfs, however, this can increase $m_\text{bol}$ by as much as 0.35 mags. 

We made the same comparisons with the 27 SEDs having $5.2\textless\lambda\textless 14.5\mu m$ spectral coverage and [3.6], [4.5], [5.8], [8], W1, W2, and W3 photometry in the MIR. Figure \ref{fig:mir_coverage}a shows how linear interpolation through IRAC+WISE photometry in lieu of an IRS spectrum produces $m_\text{bol}$ values in very good agreement with the true SED for all ultracool dwarfs to within 0.03 mags. While space based MIR spectroscopy missions such as the \textit{James Webb Space Telescope} are still vital to furthering our understanding of the atmospheres of brown dwarfs, broadband MIR photometry appears to be adequate to reconstruct the shape of ultracool dwarf SEDs at long wavelengths.

Figure \ref{fig:mir_coverage}b shows how the magnitude-magnitude relations listed in Table \ref{table:polynomials} perform instead of measured IRAC+WISE photometry. A larger scatter of about 0.1 mags is introduced in $m_\text{bol}$ as a function of spectral type though the 2nd order polynomial fit is in very good agreement with the values produced by the SEDs with MIR spectra and photometry. There are 5 L dwarf binaries that lie up to 0.2 mags below the sequence because binaries were not included in the derivation of the magnitude-magnitude relations presented in Table \ref{table:polynomials}. We also exclude them from the polynomial fits in Figures \ref{fig:opt_coverage} and \ref{fig:mir_coverage}.

The results of using a Rayleigh-Jeans tail at $\lambda\textgreater 2.5\mu m$ in the event of no available MIR data is shown in Figure \ref{fig:mir_coverage}c. The sequence crosses zero at the L/T transition but this is merely due to a coincidence of equal flux rather than the SED actually resembling a blackbody in the MIR at these spectral types. Due to the relative brightness of the K-band for late-M and L dwarfs, this approximation can overestimate the MIR flux by as much as 0.4 mags. Similarly, the relative faintness of the K-band in late-T dwarfs can underestimate $m_\text{bol}$ by as much as 0.4 mags.

\subsection{Spectral Type as a Proxy for $T_\text{eff}$}\label{sec:SpT_v_Teff}
\begin{figure*}[t]
\begin{center}
\plotone{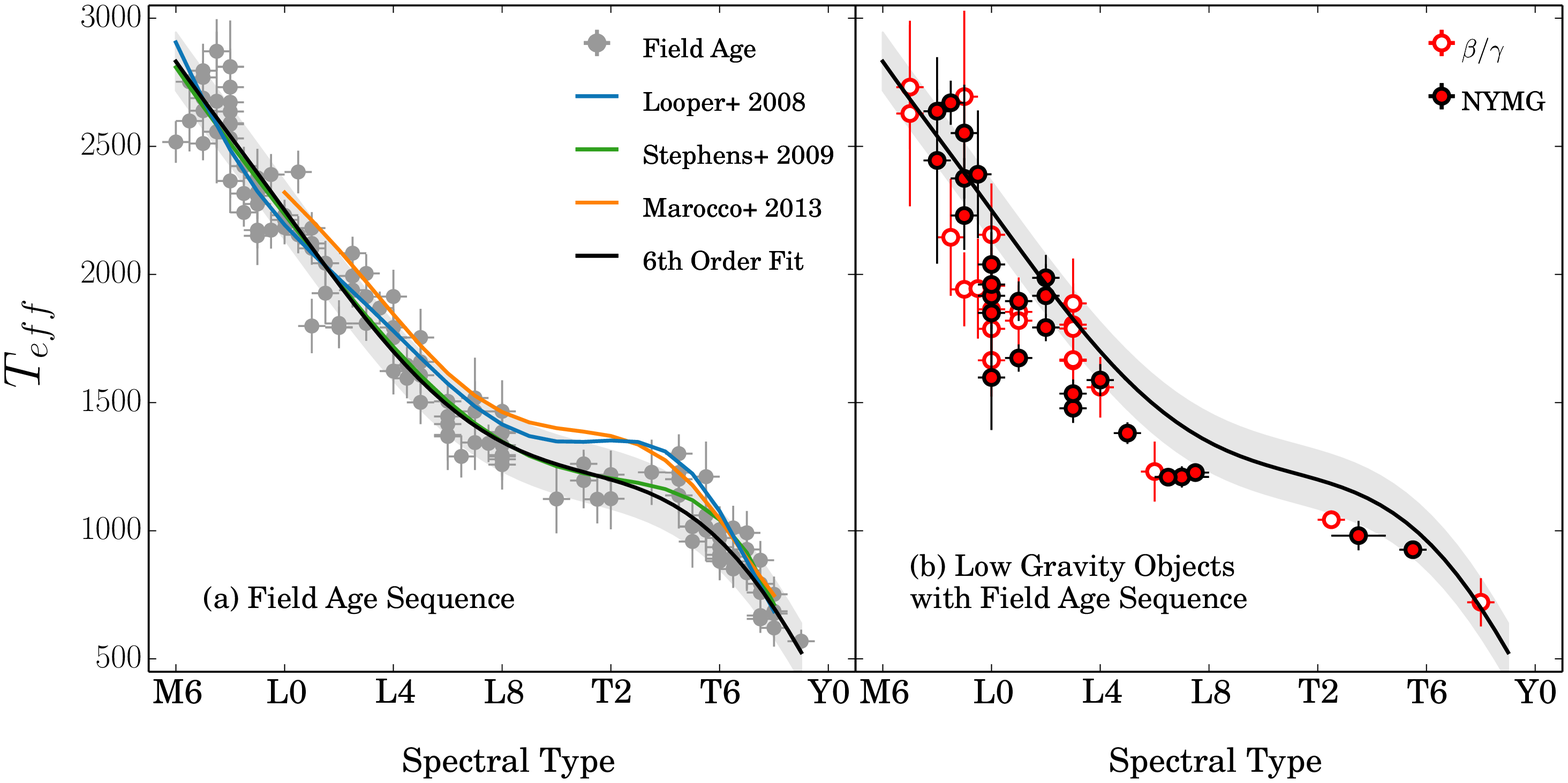}
\caption{\label{fig:spt-teff}$T_\text{eff}$ as a function of spectral type for (a) 124 field age objects and (b) 48 objects with signatures of low surface gravity and/or probable membership in a NYMG. A 6th order weighted polynomial fit (solid black line) to the field objects only is shown in both plots. The field age polynomials of \citet{Loop08b}, \citet{Step09}, and \citet{Maro13} are shown as blue, green, and orange lines, respectively.  Symbols are the same as in Figure \ref{fig:Lbol-v-SpT}.%
}
\end{center}
\end{figure*}

We use our 172 semi-empirical $T_\text{eff}$ values to revisit the temperature/spectral type relation for ultracool dwarfs. We demonstrate how most young late-M and L dwarfs lie along a cooler track than their field age counterparts of the same spectral type. This necessitates the inclusion of age and much larger uncertainties when attempting to estimate $T_\text{eff}$ from spectral type. 

We fit a 6th order polynomial weighted by the uncertainties in $T_\text{eff}$ with rms of 113K (Table \ref{table:polynomials}) to 124 field age objects in our sample (Figure \ref{fig:spt-teff}a). We find the largest outlier, the very red WISE J164715.57+563208.3 \citep[L9pec; ][]{Kirk11} which shows no signatures of youth, to be 450K cooler than the field age sequence. Also, our application of solar metallicity evolutionary models probably overpredicts the radius of the L4 subdwarf 2MASS J16262034+3925190, which would mean an even hotter and more discrepant temperature from our field age sequence. We exclude these objects from our fit due to our lack of understanding for particularly red and blue field L dwarfs. 

We also fit 6th order polynomials to 95 and 78 field age objects using their optical and IR spectral types respectively to see how $T_\text{eff}$ depended on the typing technique used. We find no significant difference in the derived polynomial for L and T dwarfs, though our late-M subsample lacks enough IR spectral types to extend the comparison earlier than M9, so we adopt optical spectral types for our M and L dwarfs and IR spectral types for the T dwarfs.

The relations of \citet{Loop08b} (blue line) and \citet{Step09} (green line) are both derived from revised versions of the \citet{Goli04a} sample with binaries removed and additional objects added. They use IR spectral types and an assumed age of 3 Gyr with uncertainties corresponding to an age range of 0.1-10 Gyr. The shape of our polynomial for field age objects (solid black line) agrees very well with the \citet{Step09} relation across most of the sequence, though we obtain temperatures about 50K cooler for mid-T dwarfs. In that work, they fit the model atmosphere grid of \citet{Saum08} to SEDs with NIR and MIR spectral coverage.

\citet{Maro13} (orange line) perform three types of atmospheric model fits to NIR spectra for their sample and calculate a weighted mean of the best fit model parameters to determine $T_\text{eff}$. In comparison to our polynomial, this produces temperatures up to 100K cooler for M7-L0 dwarfs and about 50K hotter for L3-L8 dwarfs. Their fit through the L/T transition is about 150K hotter than our field age track probably due to their unweighted fitting of only three objects in the L7-T3 range while our relation uses a weighted fit of 20 objects. All four relations agree for types later than T6.5 where the dispersion of the sequence is at a minimum of 200 K. This is most likely due to electron degeneracy putting tight constraints on predicted radii as well as atmospheric models for late-T dwarfs accurately reproducing observations \citep{delB09}.

Young objects exhibit similar or slightly higher $L_\text{bol}$ as field age objects of the same spectral type. However, their larger radii mean that young objects must have cooler photospheres than field age objects with the same $L_\text{bol}$. Figure \ref{fig:spt-teff}b shows the same field age polynomial as Figure \ref{fig:spt-teff}a with only optically typed $\beta/\gamma$ dwarfs (unfilled points with red borders) and NYMG members (red points with black borders) plotted. We find that late-M dwarfs with ages below about 25 Myr have similar temperatures as field age M dwarfs, while 25-130 Myr L0-L8 dwarfs have similar luminosities but temperatures up to 300K cooler than their field age counterparts. Also, the sensitivity of radius to the ages of young M8-L0 dwarfs creates a dispersion in $T_\text{eff}$ of 500-600 K at the M/L transition. HN Peg B (T2.5), GU Psc B (T3.5), and SDSS J111010.01+011613.1 (T5.5pec) lie about 150K below the field age T dwarf track in Figure \ref{fig:spt-teff}b but more confirmed young objects later than L8 are needed to explore this relationship in the T and Y dwarf regime.

\begin{figure}[t]
\begin{center}
\plotone{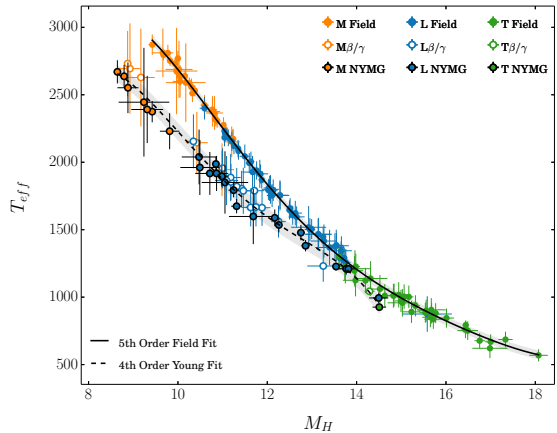}
\caption{\label{fig:H-v-teff}$T_\text{eff}$ as a function of $M_H$ for the sample. The solid black line shows the 5th order weighted polynomial fit to 115 field age objects and the dashed black line shows the 4th order weighted polynomial fit to 25 NYMG members and 21 objects with signatures of low surface gravity. The rms of each fit is shown as a grey shaded region. Young late-M to mid-L dwarfs are 400-500K cooler than field age objects of the same $M_H$ magnitude. The temperature difference in the sequences reaches a minimum of about 50K around spectral type T0. Symbols are the same as in Figure \ref{fig:Lbol-comparison}.%
}
\end{center}
\end{figure}

While spectral type is often used as a proxy for effective temperature, we find that the relationship with the smallest dispersion is $T_\text{eff}$ as a function of $M_H$ (Table \ref{table:polynomials}). Figure \ref{fig:H-v-teff} shows the 5th order weighted polynomial fit (solid black line) to 115 field age objects (grey points) with an rms of 29K. While cloud clearing exposes deeper and hotter layers causing a brightening in J-band for L/T transition objects \citep{Tinn03,Vrba04,Loop08a,Fahe12}, the monotonicity of $T_\text{eff FLD}$ with decreasing $M_H$ (and $M_{Ks}$ not shown) supports the conclusion that the opacity sources that fall below the photosphere at these spectral types are isolated to J-band \citep{Burg02b,Knap04,Fahe12}. We also fit a 4th order weighted polynomial (dashed black line) to 25 NYMG members (red points with black borders) and 21 objects with signatures of low surface gravity (unfilled points with red borders) with an rms of 60K. These sequences show that young late-M to mid-L dwarfs are about 400-500K cooler than field age objects with the same $M_H$ magnitude. Young late-L dwarfs start out about 250K cooler and then come within 50K of the field sequence at the center of the L/T transition probably due to scattering of light from H-band out to longer wavelengths by small dust grains. If a distance is available, $M_H$ can be used to more reliably estimate $T_\text{eff}$ than spectral type for both young and field age objects.

\section{Conclusions}\label{sec:conclusions}
We constructed flux calibrated $0.3-18\mu m$ SEDs using optical, NIR, and MIR spectra and photometry for 53 young and 145 field age late-M, L and T dwarfs. Gaps in data were filled by linear interpolation through photometry and newly derived age sensitive absolute magnitude-magnitude relations. These nearly complete SEDs allowed us to create a prescription for estimating systematic uncertainties in $m_\text{bol}$ of ultracool dwarfs based on the amount and quality of data available. We also presented a flux calibrated sequence of field M, L and T dwarfs for comparison with the SEDs of bona fide NYMG members.

We calculated bolometric luminosities and used evolutionary model isochrones to derive radii, effective temperatures, surface gravities, and masses for the sample, increasing the number of ultracool dwarfs with semi-empirical fundamental parameters by over 40\%. We used these new values to recalculate $L_\text{bol}$ and $T_\text{eff}$ versus spectral type relations for field age objects for comparison with young brown dwarfs. These results show that 8-130 Myr L dwarfs are up to 30$\%$ more luminous but up to 300K cooler than their field age counterparts. We also present young and field age polynomials of $T_\text{eff}$ as functions of $M_H$ that can be used estimate temperatures to within 29K and 60K respectively.

We directly compared the empirical results of this work with published $L_\text{bol}$ values inferred from model fitting and previous bolometric corrections and found disagreements of up to 30$\%$. While we quantified the suspected age dependence of temperature, magnitude, and color as functions of spectral type and luminosity, we found that our newly derived $BC_{Ks}$ is similar for late-M and L dwarfs of all ages. This correction for young objects is on average 0.05 magnitudes greater but within the rather large dispersion of the field age sequence. Our $BC_{J}$ as a function of spectral type shows two distinct tracks for young and field age objects, differing by up to a full magnitude for late-L dwarfs. 

Finally, we characterized the reddening of young L dwarfs as a shifting of flux from J-band out to W2-band, creating a pivot around K-band. Generally, this study has shown that great care must be taken when trying to determine accurate fundamental parameters of ultracool dwarfs given their ever increasing and fascinating natural diversity. 

\section*{Acknowledgements}
The authors would like to thank the anonymous referee for many helpful comments which greatly improved the manuscript. The authors would also like to thank P. A. Giorla Godfrey, A. Riedel, and M. C. Cushing for many productive discussions. This material is based upon work supported by the National Science Foundation under Grant No. 1211568 and 1313132. Support for this project was also provided by a PSC-CUNY Award, jointly funded by The Professional Staff Congress and The City University of New York, and NASA Astrophysics Data Analysis Program (ADAP) award 11-ADAP11-0169. This publication makes use of data products from: the Two Micron All Sky Survey, which is a joint project of the University of Massachusetts and the Infrared Processing and Analysis Center/California Institute of Technology, funded by the National Aeronautics and Space Administration and the National Science Foundation; the Wide-field Infrared Survey Explorer, which is a joint project of the University of California, Los Angeles, and the Jet Propulsion Laboratory/California Institute of Technology, funded by the National Aeronautics and Space Administration; the BDNYC Data Archive, an open access repository of M, L, T and Y dwarf astrometry, photometry and spectra; the SIMBAD database, Aladin, and Vizier, operated at CDS, Strasbourg, France; the M, L, T, and Y dwarf compendium housed at DwarfArchives.org; the SpeX Prism Spectral Libraries, maintained by Adam Burgasser; the Montreal Brown Dwarf and Exoplanet Spectral Library, maintained by Jonathan Gagn{\'{e}}; the NASA/ IPAC Infrared Science Archive, which is operated by the Jet Propulsion Laboratory, California Institute of Technology, under contract with the National Aeronautics and Space Administration.

\bibliography{converted_to_latex.bib}%

\clearpage 

\LongTables
\clearpage 
\begin{deluxetable*}{cccccc}
\tablecaption{The Sample of Ultracool Dwarfs with Measured SEDs \label{table:sample_table}}\tablehead{\colhead{R.A.} & \colhead{Dec} & \colhead{Designation} & \colhead{Discovery Ref} & \colhead{SpT} & \colhead{SpT Ref}}
\startdata
00 00 13.54 & $+$25 54 18.0 & SDSS J000013.54$+$255418.6 & 1 & T4.5 & 2 \\
00 04 02.88 & $-$64 10 35.8 & 2MASS J00040288$-$6410358 & 3 & L1$\gamma$ & 4 \\
00 04 34.84 & $-$40 44 05.8 & LHS 102BC & 5 & L5 & 6 \\
00 24 24.63 & $-$01 58 20.1 & 2MASS J00242463$-$0158201 & 7 & M9.5 & 8 \\
00 25 03.65 & $+$47 59 19.1 & 2MASS J00250365$+$4759191AB & 9 & L4 & 10 \\
00 27 41.97 & $+$05 03 41.7 & 2MASS J00274197$+$0503417 & 11 & M9.5$\beta$ & 8 \\
00 27 55.92 & $+$22 19 32.8 & 2MASS J00275592$+$2219328 & 12 & M8 & 13 \\
00 30 30.13 & $-$14 50 33.3 & 2MASSW J0030300$-$145033 & 14 & L7 & 14 \\
\enddata
\end{deluxetable*}
\clearpage

\begin{landscape}
\begin{deluxetable*}{ccccccccccc}
\tablecaption{Optical Photometry for the Sample \label{table:OPT_table}}\tablehead{\colhead{Designation} & \colhead{u} & \colhead{g} & \colhead{r} & \colhead{i} & \colhead{z} & \colhead{V} & \colhead{R} & \colhead{I} & \colhead{$I_{DENIS}$} & \colhead{Refs}\\ \colhead{ } & \colhead{$(0.35\mu m)$} & \colhead{$(0.48\mu m)$} & \colhead{$(0.62\mu m)$} & \colhead{$(0.76\mu m)$} & \colhead{$(0.91\mu m)$} & \colhead{$(0.54\mu m)$} & \colhead{$(0.64\mu m)$} & \colhead{$(0.80\mu m)$} & \colhead{$(0.79\mu m)$} & \colhead{ }}
\startdata
0000$+$2554 & 24.724 $\pm$ 0.927 & 25.935 $\pm$ 0.524 & 25.613 $\pm$ 0.541 & 22.771 $\pm$ 0.441 & 17.915 $\pm$ 0.031 & \nodata & \nodata & \nodata & \nodata & 1 \\
0004$-$4044BC & \nodata & \nodata & \nodata & \nodata & \nodata & 22.770 $\pm$ 0.025 & 19.238 $\pm$ 0.045 & 16.761 $\pm$ 0.012 & \nodata & 2 \\
0024$-$0158 & 21.610 $\pm$ 0.304 & 21.150 $\pm$ 0.045 & 18.517 $\pm$ 0.010 & 15.812 $\pm$ 0.005 & 13.830 $\pm$ 0.004 & 20.010 $\pm$ 0.051 & 17.448 $\pm$ 0.016 & 15.001 $\pm$ 0.025 & \nodata & 1,2 \\
0027$+$0503 & 22.421 $\pm$ 0.806 & 22.886 $\pm$ 0.181 & 21.864 $\pm$ 0.159 & 19.849 $\pm$ 0.046 & 17.990 $\pm$ 0.045 & \nodata & \nodata & \nodata & \nodata & 1 \\
0027$+$2219 & 21.273 $\pm$ 0.242 & 19.056 $\pm$ 0.010 & 17.124 $\pm$ 0.005 & 14.017 $\pm$ 0.004 & 12.175 $\pm$ 0.004 & \nodata & \nodata & \nodata & \nodata & 1 \\
0036$+$1821 & 23.778 $\pm$ 1.389 & 22.605 $\pm$ 0.137 & 19.531 $\pm$ 0.020 & 16.810 $\pm$ 0.006 & 14.677 $\pm$ 0.005 & 21.430 $\pm$ 0.024 & 18.318 $\pm$ 0.016 & 15.921 $\pm$ 0.022 & \nodata & 1,2 \\
0045$+$1634 & 23.356 $\pm$ 1.209 & 24.047 $\pm$ 0.464 & 20.107 $\pm$ 0.031 & 17.405 $\pm$ 0.008 & 15.248 $\pm$ 0.006 & \nodata & \nodata & \nodata & \nodata & 1 \\
0047$+$6803 & \nodata & \nodata & \nodata & \nodata & 18.180 $\pm$ 0.050 & \nodata & \nodata & \nodata & \nodata & 3 \\
\enddata
\tablerefs{(1) \citet{Ahn_12}; (2) \citet{Diet14}; (3) \citet{Gizi12}; (4) \citet{Epch03}; (5) \citet{Naud14}; (6) \citet{Lieb06}; (7) \citet{Chiu06}; (8) \citet{Abaz09}; (9) \citet{Legg12}; (10) \citet{Cost05}; (11) \citet{Boff14}; (12) \citet{Land09}; (13) \citet{Land92}; (14) \citet{Reyl10}; (15) \citet{Cost06}; (16) \citet{Adel08}; (17) \citet{Liu_13a}.}\tablecomments{The effective wavelength of each band is given in the table header.}\end{deluxetable*}
\clearpage
\end{landscape}

\begin{landscape}
\begin{deluxetable*}{ccccccccc}
\tablecaption{NIR Photometry for the Sample \label{table:NIR_table}}\tablehead{\colhead{Designation} & \colhead{J} & \colhead{H} & \colhead{Ks} & \colhead{$Y_{MKO}$} & \colhead{$J_{MKO}$} & \colhead{$H_{MKO}$} & \colhead{$K_{MKO}$} & \colhead{Refs}\\ \colhead{ } & \colhead{$(1.24\mu m)$} & \colhead{$(1.66\mu m)$} & \colhead{$(2.16\mu m)$} & \colhead{$(1.03\mu m)$} & \colhead{$(1.25\mu m)$} & \colhead{$(1.64\mu m)$} & \colhead{$(2.20\mu m)$} & \colhead{ }}
\startdata
0000$+$2554 & 15.063 $\pm$ 0.041 & 14.731 $\pm$ 0.074 & 14.836 $\pm$ 0.120 & 15.810 $\pm$ 0.100 & 14.730 $\pm$ 0.030 & 14.740 $\pm$ 0.070 & 14.820 $\pm$ 0.030 & 1 \\
0003$-$2822 & 13.068 $\pm$ 0.024 & 12.376 $\pm$ 0.028 & 11.972 $\pm$ 0.025 & \nodata & \nodata & \nodata & \nodata & 1 \\
0004$-$4044BC & 13.109 $\pm$ 0.024 & 12.055 $\pm$ 0.026 & 11.396 $\pm$ 0.026 & \nodata & \nodata & \nodata & \nodata & 1 \\
0004$-$6410 & 15.786 $\pm$ 0.071 & 14.831 $\pm$ 0.073 & 14.010 $\pm$ 0.045 & \nodata & \nodata & \nodata & \nodata & 1 \\
0024$-$0158 & 11.992 $\pm$ 0.035 & 11.084 $\pm$ 0.022 & 10.539 $\pm$ 0.023 & \nodata & 11.730 $\pm$ 0.030 & 11.100 $\pm$ 0.030 & 10.530 $\pm$ 0.030 & 1,2 \\
0025$+$4759 & 14.840 $\pm$ 0.038 & 13.667 $\pm$ 0.031 & 12.902 $\pm$ 0.057 & \nodata & \nodata & \nodata & \nodata & 1 \\
0027$+$0503 & 16.189 $\pm$ 0.093 & 15.288 $\pm$ 0.099 & 14.964 $\pm$ 0.116 & \nodata & \nodata & \nodata & \nodata & 1 \\
0027$+$2219 & 10.614 $\pm$ 0.022 & 9.967 $\pm$ 0.023 & 9.569 $\pm$ 0.017 & \nodata & \nodata & \nodata & \nodata & 1 \\
\enddata
\tablerefs{(1) \citet{Cutr03}; (2) \citet{Legg02a}; (3) \citet{Knap04}; (4) \citet{Legg10}; (5) \citet{Naud14}; (6) \citet{Legg98}; (7) \citet{Chiu06}; (8) \citet{Legg00}; (9) \citet{Hewe06}; (10) \citet{Legg02b}; (11) \citet{Lawr07}; (12) \citet{Curr13a}; (13) \citet{Goli04a}; (14) \citet{Burg13b}; (15) \citet{Chau04}; (16) \citet{Moha07}; (17) \citet{Burn10}; (18) \citet{Geba01}; (19) \citet{Stra99}; (20) \citet{Liu_13a}; (21) \citet{Luhm07a}.}\tablecomments{The effective wavelength of each band is given in the table header.}\end{deluxetable*}
\clearpage
\end{landscape}

\begin{landscape}
\begin{deluxetable*}{cccccccccccc}
\tablecaption{MIR Photometry for the Sample \label{table:MIR_table}}\tablehead{\colhead{\tiny Designation} & \colhead{\tiny W1} & \colhead{\tiny W2} & \colhead{\tiny W3} & \colhead{\tiny W4} & \colhead{\tiny $L_{MKO}$} & \colhead{\tiny $M_{MKO}$} & \colhead{\tiny [3.6]} & \colhead{\tiny [4.5]} & \colhead{\tiny [5.8]} & \colhead{\tiny [8]} & \colhead{\tiny Refs}\\ \colhead{ } & \colhead{\tiny $(3.35\mu m)$} & \colhead{\tiny $(4.60\mu m)$} & \colhead{\tiny $(11.56\mu m)$} & \colhead{\tiny $(22.09\mu m)$} & \colhead{\tiny $(3.75\mu m)$} & \colhead{\tiny $(4.70\mu m)$} & \colhead{\tiny $(3.51\mu m)$} & \colhead{\tiny $(4.44\mu m)$} & \colhead{\tiny $(5.63\mu m)$} & \colhead{\tiny $(7.59\mu m)$} & \colhead{\tiny  }}
\startdata
\tiny 0000$+$2554 & \tiny 14.211 $\pm$ 0.034 & \tiny 13.605 $\pm$ 0.050 & \tiny 10.282 $\pm$ 0.100 & \tiny 8.831 $\pm$ 0.503 & \tiny 13.030 $\pm$ 0.030 & \tiny 13.280 $\pm$ 0.100 & \tiny 13.720 $\pm$ 0.030 & \tiny 13.070 $\pm$ 0.030 & \tiny 12.560 $\pm$ 0.090 & \tiny 12.500 $\pm$ 0.030 & \tiny 1 \\
\tiny 0003$-$2822 & \tiny 11.671 $\pm$ 0.025 & \tiny 11.502 $\pm$ 0.023 & \tiny 10.971 $\pm$ 0.099 & \tiny \nodata & \tiny \nodata & \tiny \nodata & \tiny \nodata & \tiny \nodata & \tiny \nodata & \tiny \nodata & \tiny 1 \\
\tiny 0004$-$4044BC & \tiny 10.751 $\pm$ 0.023 & \tiny 10.488 $\pm$ 0.020 & \tiny 9.868 $\pm$ 0.045 & \tiny \nodata & \tiny \nodata & \tiny \nodata & \tiny 10.360 $\pm$ 0.010 & \tiny 10.470 $\pm$ 0.010 & \tiny 10.140 $\pm$ 0.030 & \tiny 10.130 $\pm$ 0.020 & \tiny 2,1 \\
\tiny 0004$-$6410 & \tiny 13.370 $\pm$ 0.025 & \tiny 12.937 $\pm$ 0.027 & \tiny 12.178 $\pm$ 0.244 & \tiny \nodata & \tiny \nodata & \tiny \nodata & \tiny \nodata & \tiny \nodata & \tiny \nodata & \tiny \nodata & \tiny 1 \\
\tiny 0024$-$0158 & \tiny 10.166 $\pm$ 0.024 & \tiny 9.900 $\pm$ 0.019 & \tiny 9.412 $\pm$ 0.039 & \tiny \nodata & \tiny 9.780 $\pm$ 0.130 & \tiny \nodata & \tiny 9.940 $\pm$ 0.030 & \tiny 9.910 $\pm$ 0.030 & \tiny 9.720 $\pm$ 0.010 & \tiny 9.550 $\pm$ 0.010 & \tiny 2,3,1 \\
\tiny 0025$+$4759 & \tiny 11.740 $\pm$ 0.021 & \tiny 11.572 $\pm$ 0.020 & \tiny 11.215 $\pm$ 0.085 & \tiny \nodata & \tiny \nodata & \tiny \nodata & \tiny \nodata & \tiny \nodata & \tiny \nodata & \tiny 11.485 $\pm$ 0.106 & \tiny 4,1 \\
\tiny 0027$+$0503 & \tiny 14.618 $\pm$ 0.031 & \tiny 14.071 $\pm$ 0.045 & \tiny 12.509 $\pm$ 0.529 & \tiny \nodata & \tiny \nodata & \tiny \nodata & \tiny \nodata & \tiny \nodata & \tiny \nodata & \tiny \nodata & \tiny 1 \\
\tiny 0027$+$2219 & \tiny 9.306 $\pm$ 0.022 & \tiny 9.054 $\pm$ 0.020 & \tiny 8.789 $\pm$ 0.025 & \tiny 8.653 $\pm$ 0.371 & \tiny \nodata & \tiny \nodata & \tiny \nodata & \tiny \nodata & \tiny \nodata & \tiny \nodata & \tiny 1 \\
\enddata
\tablerefs{(1) \citet{Cutr12}; (2) \citet{Patt06}; (3) \citet{Legg98}; (4) \citet{PID50367}; (5) \citet{Luhm06b}; (6) \citet{Legg02a}; (7) \citet{Legg10}; (8) \citet{Goli04a}; (9) \citet{Cutr13}; (10) \citet{Legg01}; (11) \citet{PID50059}; (12) \citet{PID3136}; (13) \citet{Kirk11}; (14) \citet{Curr13a}; (15) \citet{Legg07b}; (16) \citet{Reid02b}; (17) \citet{PID51}; (18) \citet{Burg08b}; (19) \citet{Zapa10}; (20) \citet{PID30540}; (21) \citet{PID2}; (22) \citet{Jaya03}; (23) \citet{Chau04}; (24) \citet{Grif12}; (25) \citet{Burn10}; (26) \citet{Jone96}; (27) \citet{Geba01}; (28) \citet{Dupu12a}; (29) \citet{Luhm07a}.}\tablecomments{The effective wavelength of each band is given in the table header.}\end{deluxetable*}
\clearpage
\end{landscape}

\begin{deluxetable*}{ccccccc}
\tablecaption{Optical, NIR and MIR Spectra Used to Construct SEDs \label{table:spectra_table}}\tablehead{\colhead{Designation} & \colhead{OPT} & \colhead{OPT Refs} & \colhead{NIR} & \colhead{NIR Refs} & \colhead{MIR} & \colhead{MIR Refs}}
\startdata
0000$+$2554 & \nodata & \nodata & SpeX $-$ Prism & 1 & IRS & 2 \\
0003$-$2822 & KPNO 4m $-$ R$-$C Spec & 3 & SpeX $-$ Prism & 4 & \nodata & \nodata \\
0004$-$4044BC & LRIS & 5 & SpeX $-$ Prism & 6 & IRS & 7 \\
0004$-$6410 & \nodata & \nodata & FIRE & 8 & \nodata & \nodata \\
0024$-$0158 & KPNO 4m $-$ R$-$C Spec & 3 & SpeX $-$ SXD/LXD1.9 & 9,10 & IRS & 11 \\
0025$+$4759 & \nodata & \nodata & SpeX $-$ Prism & 4 & IRS & 12 \\
0027$+$0503 & LRIS & 13 & SpeX $-$ Prism & 4 & \nodata & \nodata \\
0027$+$2219 & KPNO 4m $-$ R$-$C Spec & 14 & SpeX $-$ Prism & 4 & \nodata & \nodata \\
\enddata
\tablerefs{(1) \citet{Burg06b}; (2) \citet{Step09}; (3) \citet{Cruz07}; (4) Cruz et al. in prep; (5) \citet{Kirk00}; (6) \citet{Burg07a}; (7) \citet{PID51}; (8) Faherty et al. in prep; (9) \citet{Rayn09}; (10) \citet{Cush05}; (11) \citet{Cush06}; (12) \citet{PID50367}; (13) \citet{Leib99}; (14) \citet{Cruz03}; (15) \citet{Burg10c}; (16) \citet{Burg04c}; (17) \citet{Burg07c}; (18) \citet{Cruz09}; (19) \citet{Gizi12}; (20) \citet{Burg06c}; (21) \citet{Bard14}; (22) \citet{Scho09}; (23) \citet{Liu_13a}; (24) \citet{Kirk06}; (25) \citet{Cush09}; (26) \citet{Kirk97b}; (27) \citet{Tinn93b}; (28) \citet{Burg08d}; (29) \citet{Mace13}; (30) \citet{Kirk08}; (31) \citet{Reid08b}; (32) \citet{Chiu06}; (33) \citet{Burg08b}; (34) \citet{Kirk10}; (35) \citet{Kirk99a}; (36) \citet{PID50059}; (37) \citet{Burg03d}; (38) \citet{Cruz04}; (39) \citet{Kend03}; (40) \citet{Luyt79b}; (41) \citet{Phan03}; (42) \citet{PID3136}; (43) This work; (44) \citet{Fan_00}; (45) \citet{Goli04a}; (46) \citet{Kirk11}; (47) \citet{Bonn14b}; (48) \citet{Deac05}; (49) \citet{Loop07a}; (50) \citet{Burg10a}; (51) \citet{Alle13}; (52) \citet{Arti10}; (53) \citet{PID20409}; (54) \citet{Wils01}; (55) \citet{Ruiz01}; (56) \citet{PID20544}; (57) \citet{Fahe14}; (58) \citet{PID30540}; (59) \citet{Pati12}; (60) \citet{PID2}; (61) \citet{Gagn14b}; (62) \citet{Pati10}; (63) \citet{Burg00b}; (64) \citet{Lieb07}; (65) \citet{Loop07b}; (66) \citet{Burn11}; (67) \citet{Burg10b}; (68) \citet{Gizi00}; (69) \citet{Reid00}; (70) \citet{Kirk01a}; (71) \citet{Luyt79a}; (72) \citet{PID29}; (73) \citet{Manj14}; (74) \citet{Luhm07a}; (75) \citet{Loop08a}; (76) \citet{Loop08b}.}\end{deluxetable*}
\clearpage 

\begin{deluxetable*}{ccccc}
\tablecaption{Comparison of $L_{bol}$ Values with the Literature \label{table:Lbol_comparison_table}}\tablehead{\colhead{Designation} & \colhead{$\pi$ (mas)} & \colhead{$\log (L_*/L_\odot )$} & \colhead{$\log (L_{*,Lit}/L_\odot )$} & \colhead{Refs}}
\startdata
0004$-$4044BC & 104.7 $\pm$ 11.4 & $-$4.046 $\pm$ 0.095 & $-$4.05 $\pm$ 0.1 & 1,2 \\
0024$-$0158 & 86.6 $\pm$ 4 & $-$3.439 $\pm$ 0.040 & $-$3.505 $\pm$ 0.042 & 3,4 \\
\nodata & 86.6 $\pm$ 4 & $-$3.439 $\pm$ 0.040 & $-$3.51 $\pm$ 0.04 & 3,5 \\
0024$-$0158 & 84.2 $\pm$ 2.6 & $-$3.415 $\pm$ 0.027 & $-$3.45 $\pm$ 0.03 & 1,2 \\
0030$-$1450 & 37.42 $\pm$ 4.5 & $-$4.373 $\pm$ 0.106 & $-$4.36 $\pm$ 0.12 & 6,6 \\
0033$-$1521 & 24.8 $\pm$ 2.5 & $-$3.611 $\pm$ 0.088 & $-$3.55 $\pm$ 0.15 & 7,7 \\
0036$+$1821 & 114.2 $\pm$ 0.8 & $-$3.928 $\pm$ 0.006 & $-$3.97 $\pm$ 0.02 & 8,2 \\
\nodata & 114.2 $\pm$ 0.8 & $-$3.928 $\pm$ 0.006 & $-$3.97 $\pm$ 0.01 & 8,5 \\
\enddata
\tablerefs{(1) \citet{vanA95}; (2) \citet{Goli04a}; (3) \citet{Tinn95}; (4) \citet{Diet14}; (5) \citet{Cush05}; (6) \citet{Vrba04}; (7) \citet{Zapa14a}; (8) \citet{Dahn02}; (9) \citet{Luhm07a}; (10) \citet{Gizi15}; (11) \citet{Naud14}; (12) \citet{Tinn96}; (13) \citet{Henr06}; (14) \citet{Gate08}; (15) \citet{Liu_13a}; (16) \citet{Saum07a}; (17) \citet{Dahn02}; (18) \citet{Vrba04}; (19) \citet{Legg12}; (20) \citet{Burg08b}; (21) \citet{Tinn03}; (22) \citet{Fahe14}; (23) \citet{Mama05}; (24) \citet{Manj14}; (25) \citet{Burn11}; (26) \citet{Geba01}; (27) \citet{vanA95}; (28) \citet{vanL07}; (29) \citet{Mone92}.}\end{deluxetable*}
\clearpage 

\begin{deluxetable*}{ccccccccc}
\tablecaption{Fundamental Parameters of the Sample \label{table:params_table}}\tablehead{\colhead{Designation} & \colhead{$\pi$} & \colhead{$\pi$ Ref} & \colhead{$\log (L_*/L_\odot )$} & \colhead{Age\tablenotemark{a}} & \colhead{Radius} & \colhead{log(g)} & \colhead{$T_\text{eff}$} & \colhead{Mass}\\ \colhead{ } & \colhead{(mas)} & \colhead{ } & \colhead{ } & \colhead{(Gyr)} & \colhead{($R_{Jup}$)} & \colhead{(dex)} & \colhead{(K)} & \colhead{($M_{Jup}$)}}
\startdata
0000$+$2554 & 70.8 $\pm$ 1.9 & 1 & $-$4.68 $\pm$ 0.02 & 0.500$-$10.00 & 0.99 $\pm$ 0.15 & 5.02 $\pm$ 0.47 & 1227 $\pm$ 95 & 47.56 $\pm$ 26.76 \\
0003$-$2822 & 25.7 $\pm$ 0.93 & 2 & $-$2.95 $\pm$ 0.03 & 0.500$-$10.00 & 1.32 $\pm$ 0.05 & 5.18 $\pm$ 0.05 & 2871 $\pm$ 76 & 106.97 $\pm$ 9.15 \\
0004$-$4044BC & 77.02 $\pm$ 2.07 & 3 & $-$3.78 $\pm$ 0.02 & 0.500$-$10.00 & \nodata & \nodata & \nodata & \nodata \\
0004$-$6410 & 17 $\pm$ 1\tablenotemark{b} & 4 & $-$3.49 $\pm$ 0.06 & 0.010$-$0.040 & 1.63 $\pm$ 0.06 & 4.25 $\pm$ 0.18 & 1896 $\pm$ 77 & 19.18 $\pm$ 6.41 \\
0024$-$0158 & 86.6 $\pm$ 4 & 5 & $-$3.44 $\pm$ 0.04 & 0.500$-$10.00 & 1.09 $\pm$ 0.05 & 5.22 $\pm$ 0.10 & 2390 $\pm$ 80 & 79.27 $\pm$ 11.13 \\
0025$+$4759 & 22.8 $\pm$ 0.9 & 2 & $-$3.33 $\pm$ 0.04 & 0.500$-$10.00 & \nodata & \nodata & \nodata & \nodata \\
0027$+$0503 & 13.8 $\pm$ 1.6 & 6 & $-$3.55 $\pm$ 0.10 & 0.010$-$0.150 & 1.44 $\pm$ 0.23 & 4.48 $\pm$ 0.43 & 1945 $\pm$ 195 & 31.35 $\pm$ 19.44 \\
0027$+$2219 & 69.6 $\pm$ 0.9 & 1 & $-$2.83 $\pm$ 0.01 & 0.500$-$10.00 & \nodata & \nodata & \nodata & \nodata \\
\enddata
\tablerefs{(1) \citet{Dupu12a}; (2) \citet{vanL07}; (3) \citet{Diet14}; (4) Faherty et al. in prep; (5) \citet{Tinn95}; (6) \citet{Dahn02}; (7) \citet{Vrba04}; (8) \citet{Zapa14a}; (9) \citet{Fahe12}; (10) \citet{Gizi15}; (11) \citet{Dupu13}; (12) \citet{Maro13}; (13) \citet{Cost05}; (14) \citet{Naud14}; (15) \citet{Maro10a}; (16) \citet{Tinn96}; (17) \citet{vanA95}; (18) \citet{Patt06}; (19) \citet{Kirk12}; (20) \citet{Cost06}; (21) \citet{Mone92}; (22) \citet{Geli14}; (23) \citet{Andr11}; (24) \citet{Thor03}; (25) \citet{Legg12}; (26) \citet{Gate08}; (27) \citet{Arti10}; (28) \citet{Schi09}; (29) \citet{Burg08b}; (30) \citet{Tinn03}; (31) \citet{Luri14}; (32) \citet{Boff14}; (33) \citet{Teix08}; (34) \citet{Wein12}; (35) \citet{Duco08}; (36) \citet{Cese13}; (37) \citet{Kirk11}; (38) \citet{Reid03b}; (39) \citet{Smar13}; (40) \citet{Burg14a}; (41) \citet{Suba09}.}\tablenotetext{a}{Criteria for age range assumption is described in Section \ref{sec:radii}}.\tablenotetext{b}{Parallax calculated from kinematic distance.}\end{deluxetable*}

\clearpage
\begin{deluxetable*}{cccccccccc}\tablecaption{Polynomial Relations \label{table:polynomials}}
\tablehead{P(x)\tablenotemark{a}&x\tablenotemark{b}&rms\tablenotemark{c}&$c_0$&$c_1$&$c_2$&$c_3$&$c_4$&$c_5$&$c_6$}\startdata
$M_{r\text{ FLD}}$&9.7\textless $M_{H}$\textless 15.8&0.831&-1.209e+01&3.991e+00&-1.269e-01&1.417e-03&\nodata&\nodata&\nodata\\
$M_{r\text{ YNG}}$&8.9\textless $M_{H}$\textless 16.5&0.621&-2.487e+01&7.426e+00&-4.107e-01&8.783e-03&\nodata&\nodata&\nodata\\
$M_{i\text{ FLD}}$&9.7\textless $M_{H}$\textless 18.1&0.527&6.709e-01&2.447e-01&1.593e-01&-5.333e-03&\nodata&\nodata&\nodata\\
$M_{i\text{ YNG}}$&8.9\textless $M_{H}$\textless 16.5&0.419&-2.406e+00&1.127e+00&1.056e-01&-4.964e-03&\nodata&\nodata&\nodata\\
$M_{z\text{ FLD}}$&10.3\textless $M_{J}$\textless 18.4&0.334&-3.175e+01&7.899e+00&-4.706e-01&1.085e-02&\nodata&\nodata&\nodata\\
$M_{z\text{ YNG}}$&9.6\textless $M_{J}$\textless 16.6&0.314&-3.103e+00&1.277e+00&4.229e-02&-2.371e-03&\nodata&\nodata&\nodata\\
$M_{[3.6]\text{ FLD}}$&9.2\textless $M_{W1}$\textless 17.1&0.164&2.966e+00&3.035e-01&5.349e-02&-1.592e-03&\nodata&\nodata&\nodata\\
$M_{[3.6]\text{ YNG}}$&8.6\textless $M_{W1}$\textless 15.7&0.078&-1.776e+01&5.658e+00&-4.020e-01&1.113e-02&\nodata&\nodata&\nodata\\
\enddata
\tablenotetext{a}{$P(x) = \sum^n_{i=0} c_i x^i$}
\tablenotetext{b}{Polynomials that are a function of spectral type (SpT) accept integers 6-39 corresponding to types M6-T9. Optical spectral types are used for M and L dwarfs and infrared spectral types are used for T dwarfs. All other polynomial ranges are in magnitudes.}
\tablenotetext{c}{Units are in magnitudes except where noted.}
\end{deluxetable*}

\end{document}